\documentclass[prd,twocolumn,showpacs,amsmath,amssymb,superscriptaddress,preprintnumbers,nofootinbib]{revtex4}

\usepackage{amssymb}
\usepackage{amsmath}
\usepackage{graphicx}

\newcommand{\D}{\hat{D}}

\begin{document}
\title{Magnetic black holes and monopoles
in a nonminimal Einstein-Yang-Mills theory
with a cosmological
constant: Exact solutions}

%
\author{Alexander B. Balakin}
\email{Alexander.Balakin@kpfu.ru} \affiliation{Department of
General Relativity and Gravitation, Institute of Physics, Kazan
Federal University, Kremlevskaya str. 18, Kazan 420008, Russia}

\author{Jos\'e P. S. Lemos}
\email{joselemos@ist.utl.pt} \affiliation{Centro
Multidisciplinar de Astrof\'{\i}sica-CENTRA, Departamento de F\'{\i}sica,
Instituto Superior T\'ecnico-IST, Universidade  de
Lisboa-UL, \\ Avenida Rovisco Pais 1, 1049-001 Lisboa, Portugal}

\author{Alexei E. Zayats}
\email{Alexei.Zayats@kpfu.ru} \affiliation{Department of General
Relativity and Gravitation, Institute of Physics, Kazan Federal
University, Kremlevskaya str. 18, Kazan 420008, Russia}


\begin{abstract}

Alternative theories of gravity and their solutions are of
considerable importance since at some fundamental
level the world
can reveal new features.
Indeed, it is suspected that the gravitational field might be
nonminimally coupled to the other fields at scales
not yet probed, bringing into the forefront nonminimally coupled
theories. In this mode, we consider
a nonminimal Einstein-Yang-Mills theory with a cosmological constant.
Imposing spherical symmetry and staticity for the
spacetime and a magnetic Wu-Yang ansatz
for the Yang-Mills field, we find expressions for the
solutions of the theory. Further imposing constraints on the
nonminimal parameters, we find a family of exact solutions of the
theory depending on five parameters, namely, two
nonminimal parameters, the cosmological constant, the
magnetic charge, and the mass. These solutions represent
magnetic monopoles and black holes in magnetic monopoles
with de Sitter, Minkowskian, and anti-de Sitter
asymptotics, depending on the sign and value of the
cosmological constant $\Lambda$.
We classify completely the family of
solutions with respect to the number and the type of horizons
and show that
the spacetime solutions can have, at most, four horizons.
For particular sets of the parameters, these horizons can become
double, triple, and quadruple.
For instance, for a positive cosmological constant
$\Lambda$, there is a critical
$\Lambda_c$ for which the solution admits a quadruple
horizon, evocative of the $\Lambda_c$
that appears for a given energy density
in both the Einstein static and Eddington-Lema\^{\i}tre
dynamical universes.  As an
example of our classification, we analyze solutions in
the Drummond-Hathrell nonminimal theory that describe nonminimal black
holes.  Another application is with a set of regular black holes previously
treated.

\end{abstract}


\maketitle

\section{Introduction}

There is  great
interest in finding compact objects and black hole
solutions in all possible viable gravitational
theories, from general relativity
coupled minimally to all forms of matter, to alternative theories of
gravitation such as nonminimally coupled theories.

Vacuum spherically symmetric
general relativity contains the Schwarzschild
black hole, and when
coupled minimally to the Maxwell electromagnetic
field, contains the Reissner-Nordstr\"om black
hole, see e.g.,  \cite{griffpod}. When coupled to the
Yang-Mills field, spherically symmetric general relativity
yields
soliton \cite{bm} and
black hole solutions \cite{bizon}
with Yang-Mills hair,
and the addition of a Higgs field
produces remarkable
magnetic monopole black holes \cite{breit,lee,ortiz}.
There are many other solutions.
For instance, compact objects and
black holes appear in a non-Abelian Born-Infeld
theory
coupled to general relativity \cite{wirschins}
and in supersymmetric Einstein-Yang-Mills theories
\cite{ortin}.
In addition, regular black holes, i.e, nonsingular
black holes that
have special features, also show up
when general relativity is coupled minimally
to other fields
\cite{bardeen,beato,lemoszanchin,bbs,ma}.
Vacuum rotating stationary general relativity
contains the Kerr
black hole, and when
coupled minimally to the Maxwell electromagnetic
field, contains the Kerr-Newman black
hole, see, e.g.,  \cite{griffpod}.
Some of the spherically symmetric solutions
mentioned above also have their counterpart
when put to rotate.
To add a cosmological constant
to the general relativity equations
gives, in pure vacuum, the de Sitter (dS) solution
and the anti-de Sitter (AdS) solution,
depending on whether the cosmological constant
is positive or negative, respectively.
In such a case, the massive solutions are enlarged
to the Schwarzschild-dS solution
or the Schwarzschild-AdS solution,
depending on whether the cosmological constant
is positive or negative, respectively,
and the corresponding generalizations when one
includes electric or magnetic charge
and rotation \cite{griffpod}.
Non-Abelian monopole solutions in dS
spacetimes have been found in
\cite{v1}
and black hole hairy solutions
for a Yang-Mills field coupled
to spherically symmetric general relativity
in AdS spacetimes have also been found
\cite{wins,hosotani1,hosotani2}.

Differently from minimally coupled fields,
there are theories
that couple the gravitational field to other fields using
cross terms containing the curvature tensor.
One says then that the theory is
nonminimally coupled.
There are many fields which
can be nonminimally coupled to gravitation.
For instance, the
electromagnetic field
is nonminimally coupled
to the gravitational field in
\cite{prasanna,drumhat,balaklemos1}.

In nonminimally coupled theories
spherically symmetric solutions
with an electric field have been found in
\cite{horn1,horn2,muller}, including black holes
\cite{balakbl}
and wormholes \cite{balaklz1}.
Solutions with a magnetic field
of the Wu-Yang type in a Yang-Mills
theory nonminimally coupled to
the gravitational field have been found
as monopoles in \cite{balalz1,balakdz}, as
wormholes in
\cite{balaksz1,balaksz2}, and as regular black holes in
\cite{balaklz2}.

Here we want to proceed with the studies initiated in
\cite{balalz1,balakdz,balaksz1} and
find nonminimal solutions, now with a
generic cosmological
constant (see also \cite{balaksz2}
for wormhole solutions with a cosmological
constant). Indeed, we find a general set of
magnetic monopoles and black holes with positive, zero,
and negative cosmological constant of which the
regular black holes found in \cite{balaklz2}
are a small subset.

The nonminimal theory we use is provided in \cite{balaklemos1}, and its
extension from a Maxwell to a Yang-Mills field is in
\cite{balalz1}. In addition to the Einstein-Hilbert term and to the
Yang-Mills term, the theory
couples the Yang-Mills field
linearly to a nonminimal susceptibility tensor
in which three parameters $q_1$, $q_2$, and
$q_3$, appear as coefficients in front of the Ricci scalar, Ricci
tensor and Riemann tensor, respectively. These three parameters can
be considered phenomenological and have units of length square.

We put in the theory,
and thus in the action, a cosmological $\Lambda$ term,
which has units of inverse length square.
A positive
cosmological term appears in a cosmological framework
for providing the acceleration of the Universe
and as a setting for the dark energy. It also
appears in black hole physics. First, regular black holes
need some kind of positive cosmological constant
in its interior to provide enough repulsion that
does not allow a singularity formation
(see, e.g., \cite{lemoszanchin}).
Second, a generic cosmological constant, positive, zero,
or negative gives spacetimes
that are asympotically dS, Minkowski, or AdS.
The first case is the Universe in which we live,
the second case is mathematically simpler with
good asymptotic properties,
and
the third might be a world for elementary particles
as predicted in supergravity theories.

In considering solutions describing nonminimal
magnetic monopole stars and nonminimal
magnetic black holes we aim
to construct fully exact solutions
from the origin to infinity without the need of a
matching boundary.
Clearly it is an important task to
find continuous black hole solutions for which no
junction is needed. We
find that nonminimal coupled theories
of a Yang-Mills field with a  Wu-Yang ansatz
in spherically symmetric static spacetime yields
such solutions.
The Wu-Yang ansatz in such a
spacetime provides a magnetic
parameter $\nu$ for the Yang-Mills field,
or its square ${Q_{m}^2}=4\pi\nu^2$
giving the magnetic charge $Q_{m}$.
The
gravitational equations give the mass parameter $M$.

It is
reasonable to restrict our quest, as we do,
by imposing some asymptotic conditions at
large values of the radial variable $r$. We assume that at $r$
large one should obtain a
magnetic Reissner-Nordstr\"om-dS solution,
a  magnetic Reissner-Nordstr\"om solution,
or  a
magnetic Reissner-Nordstr\"om-AdS solution.
Generically the magnetic
Reissner-Nordstr\"om solution
with a cosmological constant is
$ds^2 = {N}(r) dt^2 - \frac{dr^2}{{N}(r)} - r^2
\left(d\theta^2 + \sin^2{\theta} d\varphi^2 \right),$ where the metric
function ${N}(r)$ is of the form
${N}(r) = 1 - \frac{2M}{r} + \frac{Q_m^2}{r^2} - \frac{\Lambda}{3} r^2$.
Here $Q_m$ is the
magnetic charge, that can be also an electric charge,
and
$M$ is the asymptotic mass of the object.
At $r\to\infty$
it has a dS asymptote, when $\Lambda >0$,
a Minkowski one, when
$\Lambda=0$, and an anti-de
Sitter asymptote, when $\Lambda<0$.

In the full nonminimal theory,
there are several characteristic scales.
The scales set by $q_1$, $q_2$, $q_3$,
the scale set by  $\Lambda$,
the magnetic charge scale $Q_{m}$,
and the mass scale $M$.
By an imposed choice we reduce the
nonminimal parameters from three
to two, having thus a five parameter
solution, with parameters
$q$, $\bar q$ say, $\Lambda$,
$Q_{m}$ and $M$.
We have then five scales
all of which are important in the
modeling of the causal structure
of the
nonminimally objects that we find,
namely,
magnetic monopoles and black holes.
We show
that in our nonminimal model the equation determining the
horizon radii can be reduced to an algebraic equation of the
sixth order and contains five
parameters.
Other radii, like those defined by minima, maxima, and
inflection are important. For instance, for a
positive cosmological constant, there is a critical
$\Lambda_c=Q_{m}^2/972 \bar q^2$ for which the solution admits a quadruple
horizon, the existence of
a $\Lambda_c$ being reminiscent of the $\Lambda_c=4\pi\rho/3$
for a given energy density
$\rho$
that characterizes both the Einstein static and Eddington-Lema\^{\i}tre
dynamical universes.
Thus, causal structure of these spacetimes is
predetermined by the interplay between the characteristic
scales.
Therefore, we
deal now with a new classification task and focus on its complete
representation.
We apply our classification to
the Drummond-Hathrell nonminimal theory that describe nonminimal black
holes and to the set of regular black holes previously
treated.

The paper is organized as follows. In Sec.~\ref{sec2}, we
revise the elements of the nonminimal
Einstein-Yang-Mills theory, the Wu-Yang-type solution for the gauge
field and write the nonminimal master equations of the model
reduced for a static
spherically symmetric spacetime.
Sec.~\ref{sec3} is devoted
to constrain the three nonminimal parameters
into two to have a five parameter family of exact
solutions of the nonminimally Einstein equations.
We present the solutions
in an explicit form, discuss the role of the
nonminimal and the other parameters
as well as putting constraining on the whole
set of parameters.
A preliminary analysis of the equations
and horizons
is performed. In
Secs.~\ref{seczclass}, \ref{secxclass}, and~\ref{sec4class}
we present the complete classification
of the solutions for
$\Lambda>0$, $\Lambda=0$ and $\Lambda<0$, respectively.
In Sec.~\ref{shortresume},
we present a table which summarized the
classification of horizons.
In Sec.~\ref{sec5},
we discuss an example of the presented
classification: we consider the causal structure of the
Drummond-Hathrell-type nonminimal theory.
In Sec.~\ref{secreg},
we mention an example of the presented
classification: the nonminimal regular
black hole.
In Sec.~\ref{Disc},
we conclude.

\section{Nonminimal Einstein-Yang-Mills theory: general formalism
and key equations for static spherically symmetric
objects}\label{sec2}

We follow the version of the nonminimal Einstein-Yang-Mills
theory, which has been formulated in
\cite{balalz1}
as an
$SU(N)$ generalization of the three-parameter nonminimal
Einstein-Maxwell theory \cite{balaklemos1}.
We recall its key elements.

\subsection{General formalism and master equations}

\subsubsection{Action functional}

The action functional for the nonminimal Einstein-Yang-Mills
theory we propose is
\begin{align}\label{act}
S_{{\rm NMEYM}} &= \int d^4 x \sqrt{-g}\
\left\{\frac{R+2\Lambda}{8\pi}+\frac{1}{2}F^{(a)}_{\,ik}
F^{ik(a)} \right.\nonumber\\
&\left.{}+\frac{1}{2}{\cal R}^{ikmn}F^{(a)}_{\,ik} F^{(a)}_{mn}
\right\}\,.
\end{align}
Here $g = {\rm det}(g_{ik})$ is the determinant of a metric tensor
$g_{ik}$, $R$ is the Ricci scalar, and $\Lambda$ is the cosmological
constant. The Einstein constant $\frac{8\pi G}{c^4}$ is reduced to
$8\pi$ as we use geometrical units, i.e., $G=1$ and $c=1$.
Latin indices
without parentheses run from 0 to 3.
$F^{(a)}_{mn}$ is the Yang-Mills
field strength, with the group index being a Latin index
with parentheses, e.g.,  $(a)$,
running from 1 to 3. Repeated group indices
should be summed with the Kronecker delta.
The nonminimal
susceptibility tensor ${\cal R}^{ikmn}$ is defined as
\begin{align}
{\cal R}^{ikmn} &\equiv \frac{q_1}{2}R\,(g^{im}g^{kn} -
g^{in}g^{km})\nonumber\\ &{}+ \frac{q_2}{2}(R^{im}g^{kn} - R^{in}g^{km} +
R^{kn}g^{im} -R^{km}g^{in})\nonumber \\ &{}+ q_3 R^{ikmn}\,, \label{sus}
\end{align}
where $R^{ik}$ and $R^{ikmn}$ are the Ricci and Riemann tensors,
respectively, and $q_1$, $q_2$, $q_3$ are phenomeno\-logi\-cal
parameters describing the nonminimal coupling of the Yang-Mills
fields with gravitation.

\subsubsection{SU(2)-symmetric Yang-Mills field}

We consider the Yang-Mills fields taking
values in the Lie algebra of the gauge group $SU(2)$,
\begin{equation}
{\bf A}_m = - i\,{\bf t}_{(a)} A^{(a)}_m  \,,
\label{represent0}
\end{equation}
\begin{equation}
{\bf F}_{mn} =
- i\,{\bf t}_{(a)} F^{(a)}_{mn} \,.
\label{represent}
\end{equation}
Here ${\bf t}_{(a)}$ are the Hermitian traceless generators of
the $SU(2)$ group, $A^{(a)}_i$ and $F^{(a)}_{mn}$ are the Yang-Mills
field potential and strength, respectively, and the group index $(a)$
runs from 1 to 3. The Yang-Mills fields $F^{(a)}_{mn}$ are
connected with the potentials of the gauge field $A^{(a)}_i$ by
the  formulas
\begin{gather}
{\bf F}_{mn} = \nabla_m {\bf A}_n - \nabla_n {\bf A}_m +
\left[{\bf A}_m,\, {\bf A}_n\right] \,,\nonumber\\
F^{(a)}_{mn} = \nabla_m A^{(a)}_n - \nabla_n A^{(a)}_m +
f_{(a)(b)(c)} A^{(b)}_m A^{(c)}_n \,. \label{Fmn}
\end{gather}
Here $\nabla _m$ is a  covariant spacetime derivative and the
symbols $f_{(a)(b)(c)}\equiv \varepsilon_{(a)(b)(c)}$ denote the
real structure constants of the gauge group $SU(2)$.

The variation of the action (\ref{act}) with respect to the Yang-Mills
potential $A_i^{(a)}$ yields
\begin{equation}
{\D}_k {\bf H}^{ik} \equiv \nabla_k {\bf H}^{ik}+\left[{\bf
A}_k,{\bf H}^{ik}\right] = 0 \,. \label{YMeq}
\end{equation}
The tensor ${\bf H}^{ik} = {\bf F}^{ik} + {\cal R}^{ikmn} {\bf
F}_{mn}$ is a non-Abelian analog of the excitation tensor, known in
electrodynamics. This analogy allows us to
consider ${\cal R}^{ikmn}$ as a susceptibility tensor.
The gauge covariant derivative $\D_m$ is defined as
\begin{equation}\label{D}
    \D_m=\nabla_m+\left[{\bf
    A}_m,\,\phantom{\bf\Phi}\right]\,.
\end{equation}

\subsubsection{Master equations for the gravitational field}

Variation of the action functional $S_{{\rm NMEYM}}$, Eq.~(\ref{act}),
with respect to
the metric $g_{ik}$ yields
\begin{equation}\label{EinMaster}
R_{ik}-\frac{1}{2}Rg_{ik}= \Lambda g_{ik}+8\pi T^{\rm
eff}_{ik}\,.
\end{equation}
The effective stress-energy tensor $T^{{\rm eff}}_{ik}$  can be
divided into four parts:
\begin{equation}
T^{\rm eff}_{ik} =  T^{YM}_{ik} + q_1 T^{I}_{ik} + q_2
T^{II}_{ik} + q_3 T^{III}_{ik} \,. \label{Tdecomp}
\end{equation}
The first term
\begin{equation}
T^{YM}_{ik} \equiv \frac{1}{4} g_{ik} F^{(a)}_{mn}F^{mn(a)} -
F^{(a)}_{in}F_{k}^{\ n(a)} \,, \label{TYM}
\end{equation}
is the stress-energy tensor of the pure Yang-Mills field. The
definitions of the other three tensors are related to the
corresponding coupling constants $q_1$, $q_2$, $q_3$.
Thus,
\begin{align}
T^{I}_{ik} &= R\,T^{YM}_{ik} -  \frac{1}{2} R_{ik}
F^{(a)}_{mn}F^{mn(a)} \nonumber\\ &{}+\frac{1}{2} \left[ {\D}_{i} {\D}_{k} -
g_{ik} {\D}^l {\D}_l \right] \left[F^{(a)}_{mn}F^{mn(a)} \right]
\,, \label{TI}
\end{align}
\begin{align}
T^{II}_{ik} &=\frac{1}{2}{\D}_l \left[
{\D}_i \left( F^{(a)}_{kn}F^{ln(a)} \right) {+} {\D}_k
\left(F^{(a)}_{in}F^{ln(a)} \right) \right] \nonumber\\
&{}-\frac{1}{2}g_{ik}\biggl[{\D}_{m}
{\D}_{l}\left(F^{mn(a)}F^{l(a)}_{\ n}\right)-R_{lm}F^{mn (a)}
F^{l(a)}_{\ n} \biggr]\nonumber \\
&{}- F^{ln(a)} \left(R_{il}F^{(a)}_{kn} +
R_{kl}F^{(a)}_{in}\right) \nonumber\\
&{}- R^{mn}F^{(a)}_{im} F_{kn}^{(a)} - \frac{1}{2} {\D}^m{\D}_m
\left(F^{(a)}_{in} F_{k}^{ \ n(a)}\right), \label{TII}
\end{align}
\begin{align}
T^{III}_{ik} &= \frac{1}{4}g_{ik}
R^{mnls}F^{(a)}_{mn}F_{ls}^{(a)}\nonumber\\
&{}- \frac{3}{4} F^{ls(a)}
\left(F_{i}^{\ n(a)} R_{knls} + F_{k}^{\
n(a)}R_{inls}\right)\nonumber \\
&{}- \frac{1}{2}{\D}_{m} {\D}_{n} \left[ F_{i}^{ \
n (a)}F_{k}^{ \ m(a)} + F_{k}^{ \ n(a)} F_{i}^{ \ m(a)} \right]
\,. \label{TIII}
\end{align}

Now we consider the formulation of the
master equations in the context of a static
spherically symmetric magnetic spacetime, with a
Wu-Yang  ansatz.

\subsection{Wu-Yang ansatz and master equations reduced to
spherical symmetry}

\subsubsection{Exact Wu-Yang magnetic-type solution
to the Yang-Mills equations}

The gauge field is considered to be characterized by the
Wu-Yang magnetic ansatz (see, e.g.,
\cite{balalz1,balakdz,balaksz1,balaksz2,balaklz2}
and references therein), i.e.,
\begin{equation}\label{1}
\mathbf{A}_{0}=\mathbf{A}_{r}=0 \,,\quad
\mathbf{A}_{\theta}=i\mathbf{t}_{\varphi},\quad
\mathbf{A}_{\varphi}=-i\nu\sin{\theta}\;\mathbf{t}_{\theta}\,.
\end{equation}
The magnetic
parameter $\nu$ is a nonvanishing integer. The generators
${\bf t}_r$, ${\bf t}_{\theta}$, and ${\bf t}_{\varphi}$ are
position-dependent and are connected with the standard
generators ${\bf t}_{(1)}$, ${\bf t}_{(2)}$, and ${\bf t}_{(3)}$
of the $SU(2)$ group as follows,
\begin{gather}
{\bf t}_r=\cos{\nu\varphi} \ \sin{\theta}\;{\bf
t}_{(1)}+\sin{\nu\varphi} \ \sin{\theta}\;{\bf
t}_{(2)}+\cos{\theta}\;{\bf t}_{(3)},\nonumber \\ {\bf
t}_{\theta}=\partial_{\theta}{\bf t}_r,\qquad {\bf
t}_{\varphi}=\frac {1}{\nu\sin{\theta}}\ \partial_{\varphi}{\bf
t}_r\,.
\label{deS5}
\end{gather}
They satisfy the following commutation relations
\begin{equation}
\left[{\bf t}_{r},{\bf t}_{\theta}\right]=i\,{\bf t}_{\varphi}
\,,\quad \left[{\bf t}_{\theta} \,, {\bf
t}_{\varphi}\right]=i\,{\bf t}_{r} \,, \quad \left[{\bf
t}_{\varphi},{\bf t}_{r}\right]=i\,{\bf t}_{\theta}\,.\label{deS6}
\end{equation}
For this ansatz, the field strength tensor
has only one nonvanishing component,
\begin{equation}\label{2}
{\bf F}_{\theta\varphi}=i\nu\sin\theta\,{\bf t}_{r}\,.
\end{equation}
Clearly, it is a magnetic-type solution and
depends essentially on the magnetic parameter $\nu$.

\subsubsection{Reduced gravity field equations}

Let us now consider a static spherically symmetric spacetime with
metric given by
\begin{equation}\label{metrica}
ds^2=\sigma^2Ndt^2-\frac{dr^2}{N}-r^2 \left( d\theta^2 +
\sin^2\theta d\varphi^2 \right) \,,
\end{equation}
where $t,r,\theta,\varphi$ spacetime spherical coordinates.
Here $\sigma$ and $N$ are functions depending on the radial
variable $r$ only. Early Einstein-Maxwell models for such a
metric with a central electric
charge \cite{horn1,horn2,muller,balakbl,balaklz1}
and a central
magnetic charge
\cite{balalz1,balakdz,balaksz1} were studied in the case
$\Lambda= 0$.
In \cite{balaksz2,balaklz2}, we eliminated this condition,
and, in particular, in \cite{balaklz2}
we studied regular black holes
with $\Lambda$ positive and negative.

The nonminimal gravity field equations in the spherical
symmetric static case have then the form
\begin{gather}
\frac{1-N}{r^2}-\frac{N'}{r} - \Lambda =
\frac{\nu^2}{r^4}\left[\frac{1}{2}-
q_1\frac{N'}{r}\right.\nonumber \\
\left.{}+(13q_1+4q_2+q_3)\frac{N}{r^2}-\frac{q_1+q_2+q_3}{r^2}\right],
\label{G1}
\end{gather}
\begin{gather}
\frac{1-N}{r^2}-\frac{N'}{r}-\frac{2N\sigma'}{r\sigma} - \Lambda =
\frac{\nu^2}{r^4}\left[\frac{1}{2}- q_1 \frac{N'}{r}- 2q_1 \frac{N
\sigma'}{r \sigma}\right.\nonumber\\ \left.{}-(7q_1+4q_2+q_3)\frac{N}{r^2}-
\frac{q_1+q_2+q_3}{r^2}\right] \,, \label{G2}
\end{gather}
\begin{gather}
\frac{1}{r}N'+N \frac{\sigma'}{r
\sigma}+\frac{3\sigma'}{2\sigma}N'+N\frac{\sigma''}{\sigma}+\frac{1}{2}
N'' + \Lambda\nonumber\\
{}= \frac{\nu^2}{r^4}\Biggl[\frac{1}{2}
-\frac{3q_1\sigma'N'}{2\sigma}-\frac{q_1N\sigma''}{\sigma}-
\frac{q_1N''}{2}\nonumber \\
{}-(7q_1+4q_2+q_3)\left(\frac{(\sigma N)'}{\sigma
r}-\frac{2N}{r^2}\right) \nonumber \\ {}+
(q_1+q_2+q_3)\frac{2}{r^2}\Biggr]\,.
\label{G3}
\end{gather}
A prime denotes a derivative with respect to the radial variable
$r$.  As usual, the compatibility conditions related to the Bianchi
identities are satisfied.

\section{Exact solutions to the gravity
field equations:  five -parameter family of exact solutions
and generic analysis}\label{sec3}

\subsection{Five-parameter family of exact solutions}\label{sec3a}

\subsubsection{General equations}

In spherical symmetry, Eq.~(\ref{G3})
is a consequence of Eqs.~(\ref{G1}) and (\ref{G2}).
The difference between Eq.~(\ref{G1}) and Eq.~(\ref{G2})
gives an equation for the function $\sigma(r)$ alone, which
does not depend on $\Lambda$, namely,
\begin{equation}
\frac{\sigma'}{\sigma} \left(1-\frac{{2 Q_{m}^2} q_1}{r^4}\right)=
\frac{{2 Q_{m}^2} (10q_1+4q_2+q_3)}{r^5}\,. \label{si}
\end{equation}
Then, Eq.~(\ref{G2}) gives the key equation for the
metric function $N(r)$,
\begin{gather}
r N' \left(1-\frac{{2 Q_{m}^2} q_1}{r^4}\right) + N \left[1 +
\frac{{2 Q_{m}^2}}{r^4} (13q_1 +4q_2 +q_3) \right] \nonumber \\ {}= 1 -
\frac{{ Q_{m}^2}}{r^2} + \frac{{2 Q_{m}^2}}{r^4} (q_1 +q_2 +q_3)-\Lambda
r^2 \,. \label{N}
\end{gather}
In Eqs.~(\ref{si}) and (\ref{N}),
\begin{equation}
{Q_{m}^2}=4\pi\nu^2\,,
\label{kappa}
\end{equation}
defining $Q_m$ as the magnetic charge
of the solution.

\subsubsection{The trivial solution}

The trivial solution in this context
is when $q_1$, $q_2$, and $q_3$ vanish,
\begin{equation}
q_1=q_2=q_3=0\,,
\label{rn10}
\end{equation}
Then in this limit, Eqs.~(\ref{si})-(\ref{N}),
admit
the solution $\sigma(r)=1$, and Eq.~(\ref{N}) yields
the Reissner-Nordstr{\"o}m solution
with a cosmological constant, i.e.,
\begin{equation}
\sigma(r)=1\,,
\label{rn110}
\end{equation}
\begin{equation}N=1-\frac{2M}{r} +
\frac{{Q_{m}^2}}{r^2} -
\frac{1}{3} \Lambda r^2
\,.
\label{rn11}
\end{equation}
It is the minimally coupled
magnetic Reissner-Nordstr\"om solution
with a cosmological constant.


\subsubsection{Solution for $q_1=0$}

 When $q_1=0$,
Eqs.~(\ref{si})-(\ref{N}) with the
asymptotic condition $\sigma(r \to \infty) \to 1$ yield
\begin{equation}
\sigma(r)=\exp\left(-\frac{Q_m^2(4q_2+q_3)}{2r^4}\right)\,,
\label{si00}
\end{equation}
\begin{align}\label{si01}
N(r)&=1-\frac{\Lambda r^2}{3}-\frac{1}{r}\,
\exp\left(\frac{Q_m^2(4q_2+q_3)}{2r^4}\right)
\nonumber\\
&{}\times\left\{2M{-}Q_{m}^2
\int\limits_r^{+\infty}\frac{dx}{x^2}\left[1-\frac{2\Lambda}{3}(4q_2+q_3)
\right.\right.
\nonumber\\
&\quad\left.\left.\quad\quad\quad{}+
\frac{{6q_2}}{x^2}\right]\exp
\left(-\frac{Q_m^2(4q_2+q_3)}{2x^4}\right)\right\}.
\end{align}
In particular, if $4q_2+q_3=0$, we have
a solution with one independent nonminimal parameter,
$q_2$ say, and so an overall four parameter solution,
given by
\begin{gather}
\sigma(r)=1,\\
N(r)=1-\frac{2M}{r}+\frac{Q_m^2}{r^2}+\frac{2Q_m^2q_2}{r^4}-\Lambda r^2.
\end{gather}
This solution has an interest of its own,
for instance, it has a more complex causal structure
than the Reissner-Nordstr\"om solution.
But here we want to discuss the more general case
when $q_1\neq0$ giving a solution which
in general
has two independent nonminimal parameters,
and so it is a five parameter solution

\subsubsection{General solution}

To find the general solution
we define $\xi$ as
\begin{equation}
\xi \equiv \frac{10q_1+4q_2+q_3}{4q_1} \,.
\label{si2xi}
\end{equation}
Then, for generic $q_1$,
$q_2$, $q_3$,  Eq.~(\ref{si})
together with the
asymptotic condition $\sigma(r \to \infty) \to 1$ yields
\begin{equation}
\sigma(r)=\left(1-\frac{{2Q_{m}^2} q_1}{r^4}\right)^{\xi}\,.
\label{si2}
\end{equation}
The cases $q_1=q_2=q_3=0$, and
$q_1=0$ and $4q_2+q_3=0$ are particular cases
of Eq.~(\ref{si}) which we have mentioned.
Eq.~(\ref{N}) together with the
asymptotic condition  $N(r
\to \infty) \to 1- \Lambda r^2/3$ yields
\begin{align}\label{si23}
N&=1-\frac{\Lambda r^2}{3}-\frac{1}{r}\,\left(1-\frac{{2Q_{m}^2}
q_1}{r^4}\right)^{-(\xi+1)}
\nonumber\\
&{}\times\left\{2M{-}
\int\limits_r^{+\infty}\frac{dx}{x^2}\left[Q_{m}^2
\left[1-\frac{2\Lambda}{3}(11q_1+4q_2+q_3)\right]
\right.\right.
\nonumber\\
&\quad\left.\left.\quad\quad\quad{+}
\frac{{6Q_{m}^2}}{x^2}(4q_1{+}q_2)\right]\left(1{-}\frac{{2Q_{m}^2}
q_1}{x^4}\right)^{\xi}\right\},
\end{align}

The general setup
provided by Eqs.~(\ref{si2xi})-(\ref{si23})
yields, so far, a six-parameter family of exact
solutions: the three nonminimal
parameters $q_1$, $q_2$, and $q_3$,
the cosmological constant $\Lambda$,
the magnetic charge
$Q_{m}$ of the Wu-Yang gauge field,
and
the mass $M$
of the solution.

\subsubsection{General analysis}

Equations~(\ref{si2xi})-(\ref{si23})
yield a six-parameter family of exact
solutions. The parameters are
the three nonminimal
parameters $q_1$, $q_2$, and $q_3$,
the cosmological constant $\Lambda$, the magnetic charge parameter
$Q_{m}$, and the asymptotic mass $M$.
We now choose one appropriate
relation between the
 three nonminimal
parameters $q_1$, $q_2$, and $q_3$
and reduce the family to a five-parameter family of exact
solutions.
For that we have to discuss the parameter $\xi$
given in Eq.~(\ref{si2xi}).
To find constraints on the parameter $\xi$, and so, on
$q_1$, $q_2$, and $q_3$, we study
the behavior of the functions at
some finite $r$, $r \to \infty$,
and $r \to 0$.

For some finite $r$,
we find that we should put
$q_1\leq 0$. This is because
for some finite $r$,
when $q_1>0$ nasty singularities appear in  $\sigma(r)$
and $N(r)$ in
Eqs.~(\ref{si2})-(\ref{si23}).
The case $q_1=0$ was treated before
and it has curvature singularities at $r=0$.
Thus we put
\begin{equation}\label{q1<0}
q_1<0\,.
\end{equation}

When $r \to \infty$, these solutions, Eqs.~(\ref{si2})-(\ref{si23}),
asymptotically behave as
\begin{align}
\sigma&=1 + \frac{{2Q_{m}^2} q_1}{r^4}\,\xi + \dots\,,\\
N&=-\frac{\Lambda r^2}{3}+1-\frac{2M}{r}
\nonumber\\
&+\frac{Q_{m}^2}{r^2}
\left[1-\frac{2\Lambda}{3}(11q_1+4q_2+q_3)\right]
\nonumber\\
&+\frac{{2Q_{m}^2}}{r^4}\, (4q_1+q_2)+ \dots \,.
\label{N26}
\end{align}
Thus, $\sigma(\infty)=1$ for arbitrary $\xi$. As for $N(r \to
\infty)$, it  displays a dS asymptotic behavior
when $\Lambda$ is
positive, a
Minkowski asymptotic behavior
when $\Lambda=0$, and
an AdS
behavior when $\Lambda$ is negative. 
Thus, there are no constraints on
the parameter $\xi$ in the limit $r \to \infty$.

When $r \to 0$ the analysis is subtle.
It should be divided into two cases,
$\xi< -3/4$ and $\xi\geq -3/4$.

\noindent $\xi< -3/4$:
\noindent
When $\xi\leq -3/4$ and $q_1<0$, we have $\sigma(0) =
0$ and $N(0) = \infty$; when $q_1=0$ and $4q_2+q_3>0$,
the metric functions have the same behavior at
the origin; if $q_1=0$ and $4q_2+q_3<0$, $\sigma(0) =
\infty$ and $N(0)$ is finite. Finally, if $q_1=0$, $4q_2+q_3=0$,
but $q_2\neq0$, we have $\sigma(0) =
1$ and $N(0) = \infty$.  From the point of view of invariants
divergency,
all these cases blow up at the origin faster
than $1/r^4$.

\noindent $\xi\geq -3/4$:
\noindent
When $\xi\geq-3/4$ and $q_1=0$ curvature
singularities appear at the origin.
On the
other hand,
when $\xi\geq-3/4$ and $q_1<0$, the Ricci scalar square,
the Ricci tensor square scalar, and
the Kretschmann scalar are given as $r\to 0$
by,
\begin{equation}
R^2=\frac{4}{r^4}\,\left[4\xi N(0)(4\xi-1)+N(0)-1\right]^2 \,.
\label{rscalar}
\end{equation}
\begin{eqnarray}
&R_{ik}R^{ik}=
\frac{2}{r^4}\left[32\xi^2 N(0)^2(8\xi^2+1)-
\nonumber\right. \\
&
\left. 8 \xi N(0)
(N(0)-1)+(N(0)-1)^2\right]\,,
\label{rsquarescalar}
\end{eqnarray}
\begin{eqnarray}
&R_{ikmn}R^{ikmn}=\frac{4}{r^4}\left[16\xi^2
N(0)^2(16\xi^2+8\xi+3)+\nonumber\right. \\
&
\left. (N(0)-1)^2\right]\,,
\label{kscalar}
\end{eqnarray}
respectively, and
where $N(0)$ is $N(r)$ at $r=0$, i.e.,
\begin{equation}
N(0) = 1- \frac{3(4q_1+q_2)}{q_1(4\xi+3)}
= \frac{q_1+q_2+q_3}{13q_1+4q_2+q_3} \,.
\label{nof0}
\end{equation}
So all  quadratic curvature invariants behave at $r \to 0$
according to the formula
\begin{equation}
{\rm Inv(r\to0)}= \frac{C}{r^{4}}\,,
\label{inv}
\end{equation}
for some constant $C$ that can be extracted
from Eqs.~(\ref{rscalar})-(\ref{nof0}).
We find that $C\geq0$. The case $C=0$ happens when
$N(0)=1$ and $ \xi = 0 $. In this case,
all invariants are zero at the center everywhere
and the corresponding solutions yield regular
objects and regular black holes. This case has
been treated in \cite{balaklz2}.
Although important this case is too particularized.
In order to pick up a more general
and also interesting case for $\xi>-3/4$ and $q_1<0$,
we have to address the
behavior of the functions $\sigma(r)$ and $N(r)$.
(a) The case $N(0)$ finite
(and not equal to 1) and $\xi = 0$
stands out clearly. In this case, $\sigma(0) =
1$,
$g_{00}(0) = N(0)$ is finite and  $g_{rr}(0)  = 1/N(0)$ is
also finite. Thus, the singularities that appear
at $r=0$ are of the conical
type and so are milder singularities.
To be complete let us list the other cases:
(b)
when $N(0)=1 $ and $-3/4 <\xi < 0$, then $\sigma(0) = 0$,
$g_{00}(0)=0$, $g_{rr}(0) \neq \infty$;
(c) when $N(0) \neq 1$, $N(0) \neq 0$, but $-3/4 <\xi < 0$, here
$\sigma(0) = 0$ and $g_{00}(0)=0$, $g_{rr}(0) \neq \infty$;
(d)
 when $N(0) = 0$, and $-3/4 <\xi < 0$, here $\sigma(0) = 0$ and
$g_{00}(0)=0$, $g_{rr}(0) = \infty$.
We opt for studying case (a) in detail.

So, below we consider
models with $\xi=0$, or $10q_1+4q_2+q_3=0$, and $q_1<0$ only,
thus reducing the six-parameter family of
solutions to a five-parameter one.
This five-parameter family of
solutions has many particular cases.
The trivial case $q_1=q_2=q_3=0$, and the case $q_1=0$
and $q_2$ and $q_3$ free have already been treated
and will not take part in the following analysis.
The case $10q_1+4q_2+q_3=0$
and $4q_1+q_2=0$ is the regular black hole case
already mention and studied in detail in
\cite{balaklz2}. And, of course, there are other examples
as we will see below.

So let us consider explicitly the condition
$\xi=0$, i.e.,
\begin{equation}
10q_1+4q_2+q_3=0 \,,
\label{qsrel}
\end{equation}
which guarantees that $\sigma(r) = 1$.
We define $q$ and $\bar q$ such that
\begin{equation}
q \equiv
-q_1 \,,
\label{q1}
\end{equation}
\begin{equation}
\bar{q} \equiv q_2 +3q_1  \,,
\label{q2}
\end{equation}
so that from Eq.~(\ref{qsrel}),
we find
\begin{equation}
 - 2q - 4\bar{q}= q_3 \,.
\label{q3}
\end{equation}
Due to Eq.~(\ref{qsrel}), we have now two
nonminimal parameters, $q$ and $\bar q$,
instead of the initial three.
From Eq.~(\ref{q1<0}), assume
$
q>0$,
so that there are no wild singularities
at finite $r$.

\subsubsection{Explicit five-parameter family of solutions
with two nonminimal parameters $q$ and $\bar q$,
$\Lambda$, $Q_{m}$, and
$M$}\label{sec3b}

We thus deal with a
five-parameter family of solutions,
$q$, $\bar{q}$, $\Lambda$, $Q_{m}$, and $M$,
instead of six.
The metric
functions $\sigma(r)$ and $N(r)$ take then
the following explicit form
\begin{equation}
\sigma(r) \equiv 1\,,
\label{sigma1}
\end{equation}
\begin{equation}
N=1-\frac{\Lambda r^2}{3}+\frac{r^2 {Q_{m}^2}
\left(1+\frac{2\Lambda q}{3}\right)-2Mr^3 + {2Q_{m}^2}
(\bar{q}-q)}{r^4+{2Q_{m}^2} q} \,,
\label{N00}
\end{equation}
When $q > 0$, this function is finite for all finite values of
$r$, and the value $N(0)$ is equal now to $N(0) =
\frac{\bar{q}}{q}$.  The first derivative
takes zero value at the center, $N'(0)=0$. The second
derivative at the center $N''(0) = \frac{1}{q}$ depends on the
nonminimal coupling parameter $q$ only, and is positive for $q>0$.
This means that $r=0$ is the minimum of the regular
function $N(r)$, which near the center has the form
\begin{equation}\label{N002}
N(r) =  \frac{\bar{q}}{q} + \frac{1}{2 q} r^2 + ...\,.
\end{equation}
In addition to the root $r=0$, the equation $N'(r)
{=}0$ can have other root $r=r_{{\rm min}}>0$ related to a
minimum of the function $N(r)$. Clearly, at this radius
any massive particle can be in
a stable equilibrium.
This point is a finite $r$ equilibrium point in contrast
to the central equilibrium point $r=0$.

\subsubsection{Roles and constraints on the five
parameters
 $q$ and $\bar q$,
$\Lambda$, $Q_{m}$, and
$M$}

We now discuss the roles and constraints on the
five parameters $q$, $\bar{q}$, $\Lambda$, $Q_{m}$, and $M$.
As we have pointed out the parameter $q$ must obey
\begin{equation}
q>0 \,,
\label{q>0}
\end{equation}
such that there are no singularities at some finite $r$.
The parameter  ${\bar q}$ should obey
\begin{equation}
-\infty <{\bar q}<\infty  \,,
\label{barq>0}
\end{equation}
i.e., it is not restricted.

The main role of the
parameter $\Lambda$ is at infinity. At $r=0$ the
cosmological term $\Lambda r^2$ vanishes.
At infinity one has a spacetime asymptotically
dS for
\begin{equation}
\Lambda>0 \,,
\label{l>0}
\end{equation}
asymptotically
Minkowski for
\begin{equation}
\Lambda=0 \,,
\label{l=0}
\end{equation}
and
asymptotically
AdS for $\Lambda<0$.
From Eq.~(\ref{N00}),
we see that in order to have
a solution with $Q_{{\rm eff}\,m}^2\equiv{Q_{m}^2}
\left(1+\frac{2\Lambda q}{3}\right)\geq0$, we must impose
\begin{equation}
-\frac{3}{2q}\leq\Lambda<0 \,.
\label{l<0}
\end{equation}

The parameter ${Q_{m}^2} \equiv 4\pi \nu^2$ is  the
magnetic charge of the Wu-Yang field.
Not wanting to consider imaginary
 Wu-Yang magnetic charge  $\nu$ we discuss solutions for
which
\begin{equation}
{Q_{m}^2}>0 \,.
\label{kappa>0}
\end{equation}
Moreover, the redefined nonminimal coupling
constants $q$ and  $\bar q$ enter the solutions
given in Eq.~(\ref{N00})
in the form of a product with
${Q_{m}^2}$. This means that the case $Q_{m}=0$
does not yield
nonminimal solutions.
In addition, one sees that indeed the product $({Q_{m}^2} q)^{\frac14}$
plays the role of an effective
nonminimal scale (see, Eq.~(\ref{N00})), and
the value of $({Q_{m}^2} q)^{\frac14}$ predetermines the
number and type of horizons.

The parameter $M$ is the asymptotic mass.
In order to have the usual solutions with positive mass at infinity,
we impose
\begin{equation}
M>0 \,.
\label{M>0}
\end{equation}

\subsection{Horizon classification: Auxiliary
function and preliminary analysis}
\label{sec3c}

When  Eq.~(\ref{N00}) obeys
\begin{equation}
N(r_{h})=0\,,
\label{rs}
\end{equation}
and the roots $r_{h}$ are real and
positive, we are in the presence
of horizons at those radii.
Since the equation $N(r_{h})=0$ can be reduced to an algebraic
equation of order six, the number of
horizons is not more than six. We show that, in fact, the
number of horizons cannot be more than four.

The horizon number is of major importance
in the study of any spacetime causal
structure. In this structure it is also important
to classify the horizons and visualize them in
figures. In order to classify the solutions with a different number
of horizons, we used here a method applied earlier
\cite{balakdz} (see also \cite{balaklz2}). This method is based on the
introduction of an auxiliary function $f(r)$ in the following
context. The equation $N(r)=0$ with $N(r)$
given in Eq.~(\ref{N00}) can be
rewritten in the form
\begin{gather}
2M = f(r) \,,\label{revN00}\\
f(r) \equiv  -\frac{\Lambda r^3}{3} + r +
\frac{{Q_{m}^2}}{r} + \frac{{2Q_{m}^2} \bar{q}}{r^3} \,.
\label{revN00xx}
\end{gather}
To count the horizon number we have to determine the number of
points in which the plot of the function $y=f(r)$ is crossed by
the horizontal mass line $y=2M$. From a physical point of
view, this procedure shows how many horizons there are, when
the mass of the object is equal to $M$.
Clearly,
the parameter $\bar{q}$
regulates the behavior of $f(r)$ at $r\to 0$,
and
the parameter $\Lambda$ predetermines the
behavior of $f(r)$ at $r\to \infty$. Thus, these two
parameters are the principal parameters in this analysis. Below
we describe the details of the corresponding classification.

To proceed we have to analyze the equation
$f'(r)=0$
which
gives the extrema of the function $f(r)$.
The equation
$f'(r)=0$ can be
rewritten as the bicubic equation
\begin{equation}\label{revN010}
\Lambda r^6 -r^4 + {Q_{m}^2} r^2 + 6 {Q_{m}^2} \bar{q} =0 \,.
\end{equation}

We also have to analyze the important equation $f''(r)=0$
which can also be rewritten as the following bicubic equation
\begin{equation}\label{revN01}
-\Lambda r^6 + {Q_{m}^2} r^2 + 12{Q_{m}^2} \bar{q} =0 \,.
\end{equation}

First we analyze the
$f'(r)=0$ equation.
Considering then the bicubic equation,
Eq.~(\ref{revN010}),
in terms of the auxiliary quantity,
\begin{equation}
X \equiv r^2-\frac{1}{3\Lambda}\,,
\label{Xr2}
\end{equation}
one gets a cubic equation of the form
\begin{equation}
X^3 + L X +M =0\,.
\label{X3+}
\end{equation}
The corresponding discriminant ${D}=-
(4{{L}}^3+27{{M}}^2)$ is now of the form,
\begin{eqnarray}
{D}=& -\frac{3{Q_{m}^4}}{2\Lambda^4}
\left\{\left(1+ 18 \Lambda \bar{q} + \frac{{Q_{m}^2}}{27
\bar{q}} \right)^2\right. \nonumber\\
&\left. - \left[\left(1+ \frac{{2Q_{m}^2}}{54
\bar{q}}\right)^2 + \frac13
+ 8 \frac{\bar{q}}{{Q_{m}^2}} \right] \right\}\,.
\label{Xr23}
\end{eqnarray}
Equation~(\ref{X3+}) gives three roots $X_{1}$,
$X_{2}$, $X_{3}$,
with properties
\begin{equation}
X_{1} X_{2} X_{3} = - \frac{6{Q_{m}^2} \bar{q}}{\Lambda} \,,
\quad X_{1} + X_{2} +  X_{3} = \frac{1}{\Lambda} \,.
\label{revN051}
\end{equation}
Clearly, when $\bar{q}$ is negative
and $\Lambda$ is positive, it
is  admissible that all three roots are positive, thus,
three is the maximal number of extrema.
When ${D}=0$, we deal with a special value
of the parameter, ${\Lambda}_{{\rm special}}$, given by
\begin{equation}
{\Lambda}_{{\rm special}}= \frac{1}{18\bar{q}}
\left[- \left(1+\frac{{Q_{m}^2}}{27 \bar{q}} \right)
\pm \sqrt{\left(1+\frac{{Q_{m}^2}}{18 \bar{q}} \right)^3
\frac{8\bar{q}}{{Q_{m}^2}}} \right] \,.
\label{Xr25}
\end{equation}
When $\bar{q}\geq -\frac{{Q_{m}^2}}{18}$,
${\Lambda}_{{\rm special}}$ is real.
When
$\bar{q} = -\frac{{Q_{m}^2}}{18}$, we have that
${\Lambda}_{{\rm special}}=\frac{1}{3{Q_{m}^2}}$, which is positive.

We also have to
analyze the important equation $f''(r)=0$, i.e.,
Eq.~(\ref{revN01}).
This equation gives the inflexion points of the function
$f(r)$. The analysis on the number
of inflexion points, provided by Eq.~(\ref{revN01}) is a
convenient tool for the required classification.
The analysis
of extrema, derived from Eq.~(\ref{revN010}),
is  a supplementary tool describing some details of the
classification.
In terms of the auxiliary variable,
\begin{equation}
Y \equiv r^2 \,,
\label{Yr2}
\end{equation}
Eq.~(\ref{revN01})
gives a cubic equation of the type
\begin{equation}
Y^3 + PY +Q
=0\,,
\label{Y3+}
\end{equation}
for which the discriminant is
$\Delta \equiv - (4P^3+27Q^2)$, i.e.,
\begin{equation}\label{revN04}
\Delta  = \frac{{Q_{m}^6}}{ \Lambda^2}\left(
\frac{1}{\Lambda} - \frac{972}{{Q_{m}^2}} {\bar{q}}^2 \right) \,.
\end{equation}
The product and the sum of the roots of Eq.~(\ref{Y3+})
are, respectively,
\begin{equation}\label{revN05}
Y_{1} Y_{2} Y_{3} = \frac{12{Q_{m}^2} \bar{q}}{\Lambda}
 \,, \quad Y_{1} + Y_{2} +  Y_{3} =0 \,.
\end{equation}
Since the sum of roots is equal to zero, at least one real root is
negative or the real part of complex conjugated pair of roots is
negative. Thus, two is the maximal number of inflexion
points.
Since the sign of
the product $Y_{1} Y_{2} Y_{3}$ of the roots depends on the
ratio of the parameters $\bar{q}$, $\Lambda$,
and $Q_{m}^2$, (see
Eq.~(\ref{revN05})),
for the
$\Lambda>0$ case,
below we distinguish eight
different situations using
the critical value of the cosmological constant $\Lambda_{c}$,
for which the discriminant $\Delta$ vanishes, i.e.,
\begin{equation}
\Lambda_{c} = \frac{{Q_{m}^2}}{972\,{\bar q}^2}\,.
\label{lambdac}
\end{equation}
So, clearly, $\Lambda_{c}>0$.
The existence of
a $\Lambda_c$ is reminiscent of the $\Lambda_c=4\pi\rho/3$
for a given energy density
$\rho$
that characterizes both the Einstein static and Eddington-Lema\^{\i}tre
dynamical universes.

According to these preliminary considerations, the classification of
horizons can be based on the analysis of the
Eqs.~(\ref{revN00})-(\ref{revN00xx})
with respect to the parameters  $\bar{q}$
and $\Lambda$. Below, we fix the cosmological
constant according to $\Lambda>0$,
$\Lambda=0$, and
$\Lambda<0$, and vary the
parameter $\bar{q}$.

\section{Exact solutions
With a positive
cosmological constant, $\Lambda >0$}
\label{seczclass}

\begin{figure*}[t]
\halign{\hfil#\hfil&\;\hfil#\hfil&\;\hfil#\hfil&\;\hfil#\hfil\cr
\includegraphics[height=3.35cm]{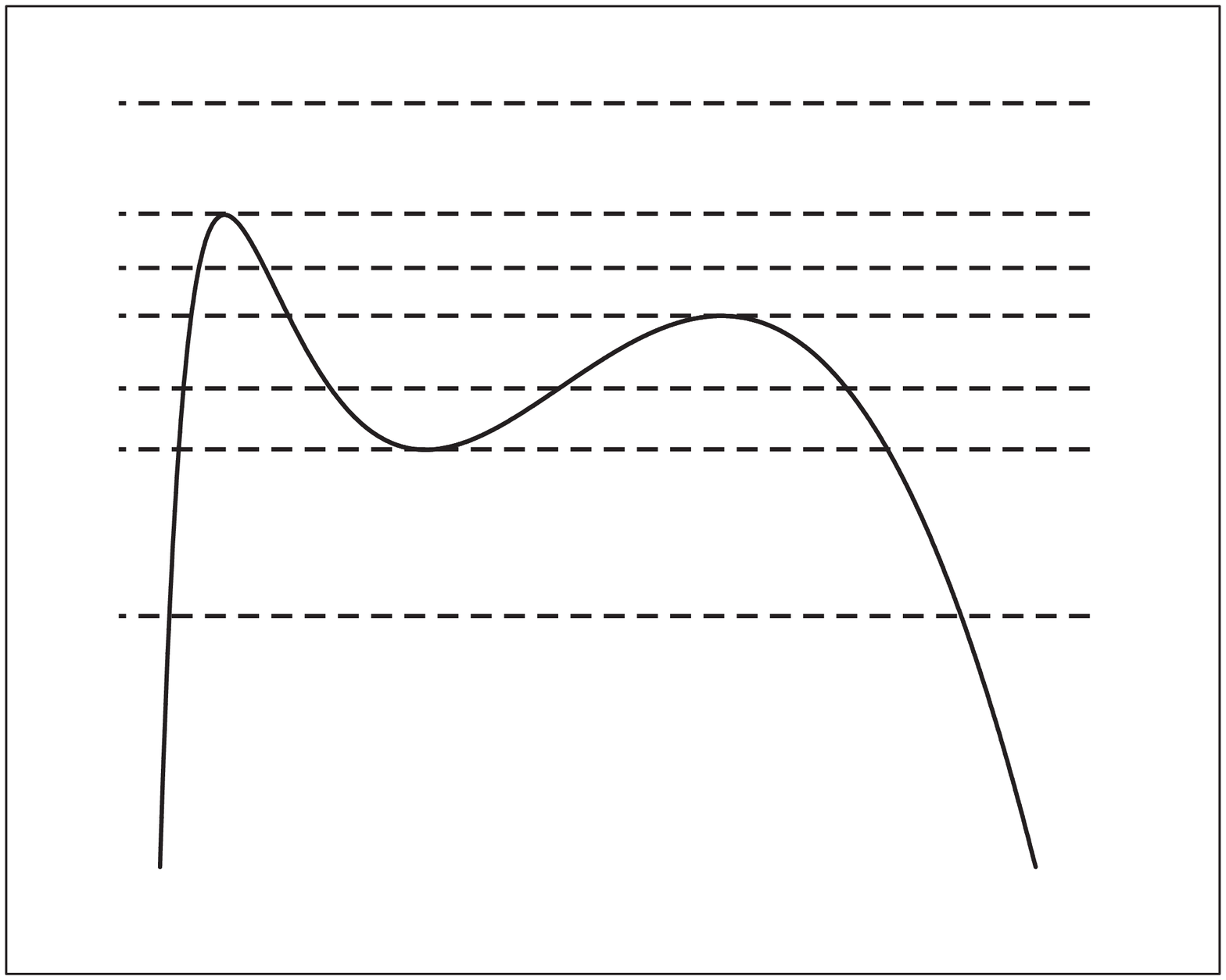}
&\includegraphics[height=3.35cm]{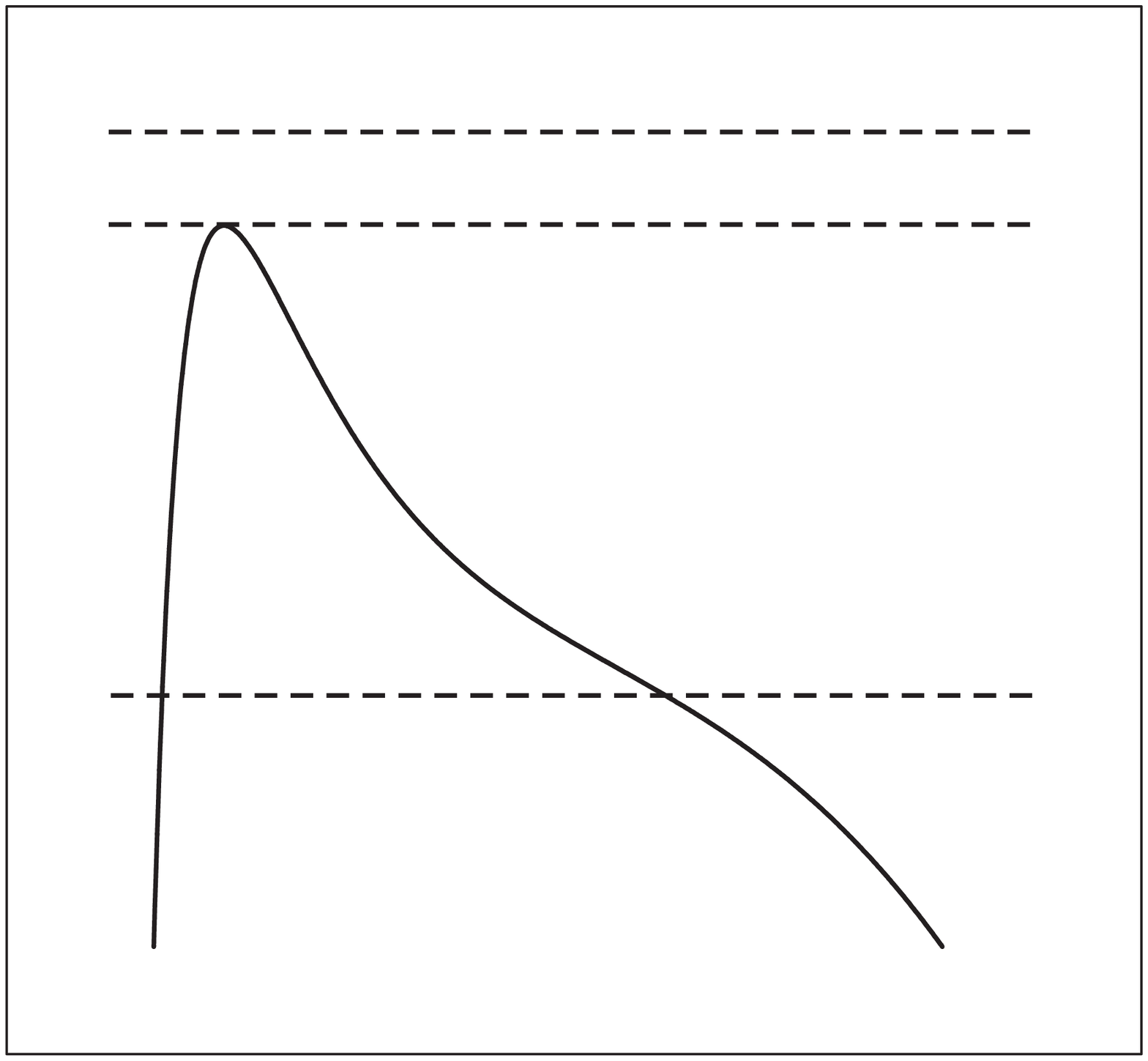}&
\includegraphics[height=3.35cm]{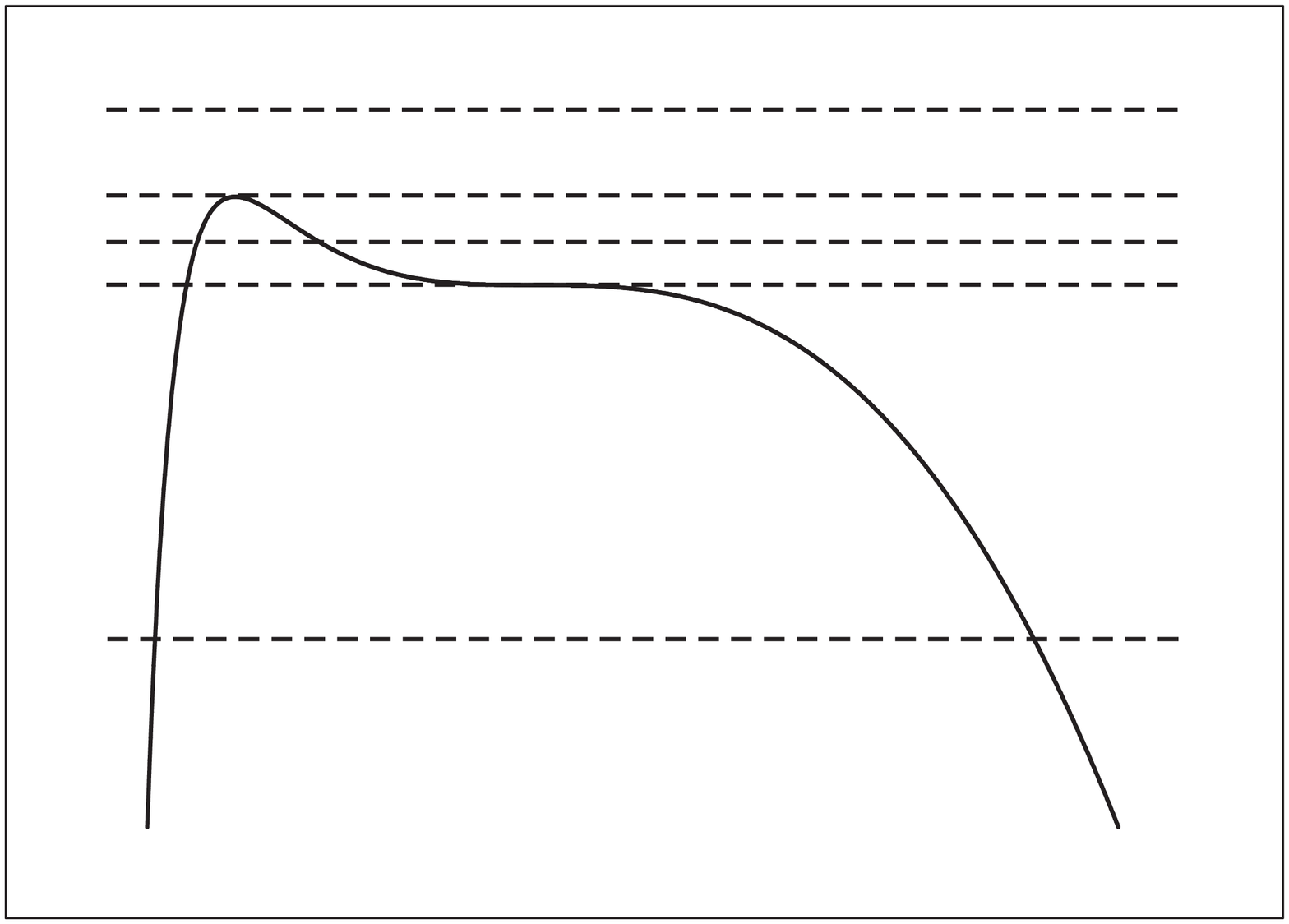}&
\includegraphics[height=3.35cm]{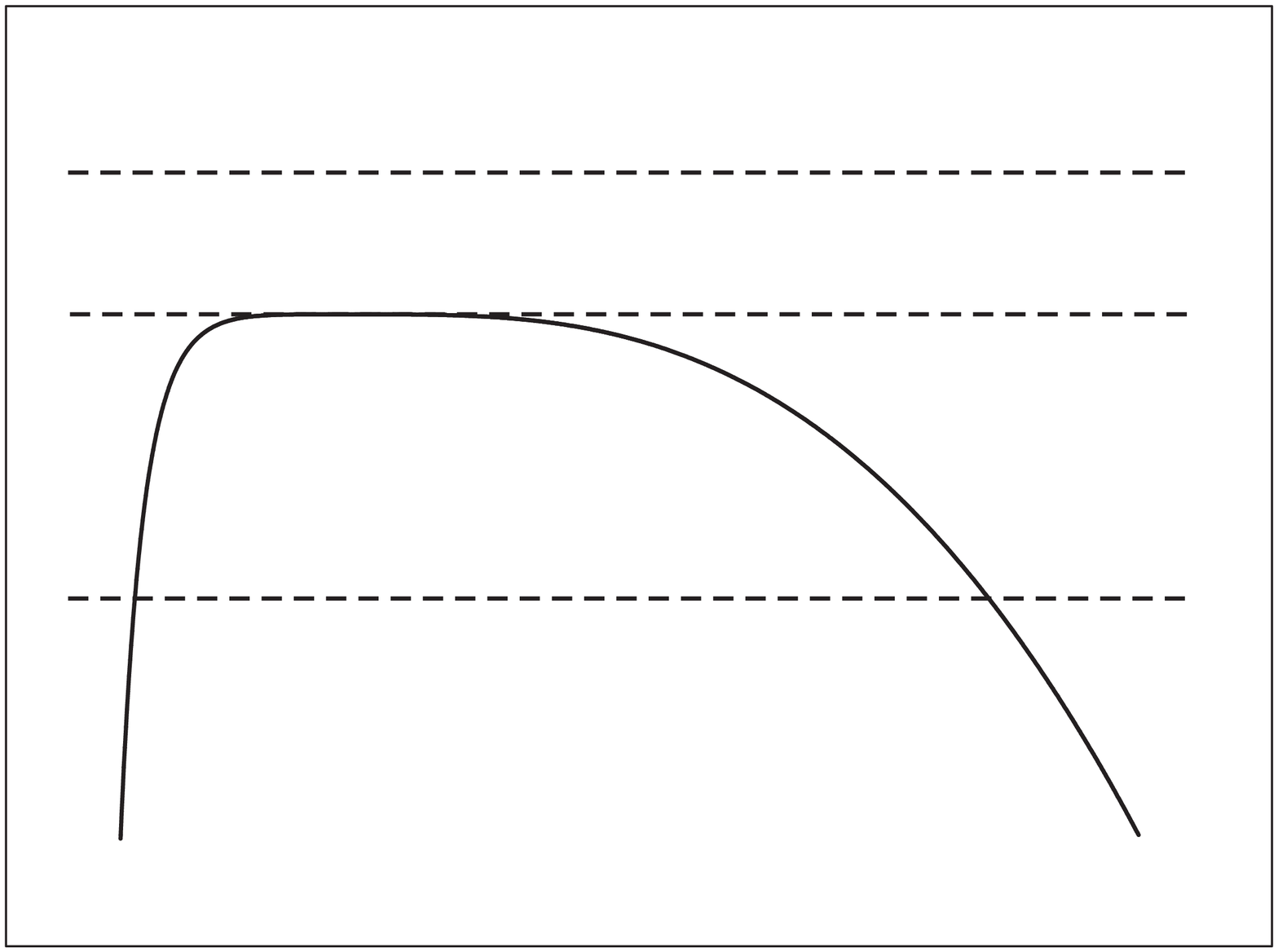}\cr
$(1a)$&$(1b)$&$(1c)$&$(1d)$\cr}

\medbreak

\halign{\hfil#\hfil&\;\hfil#\hfil&\;\hfil#\hfil&\;\hfil#\hfil\cr
\includegraphics[height=3.35cm]{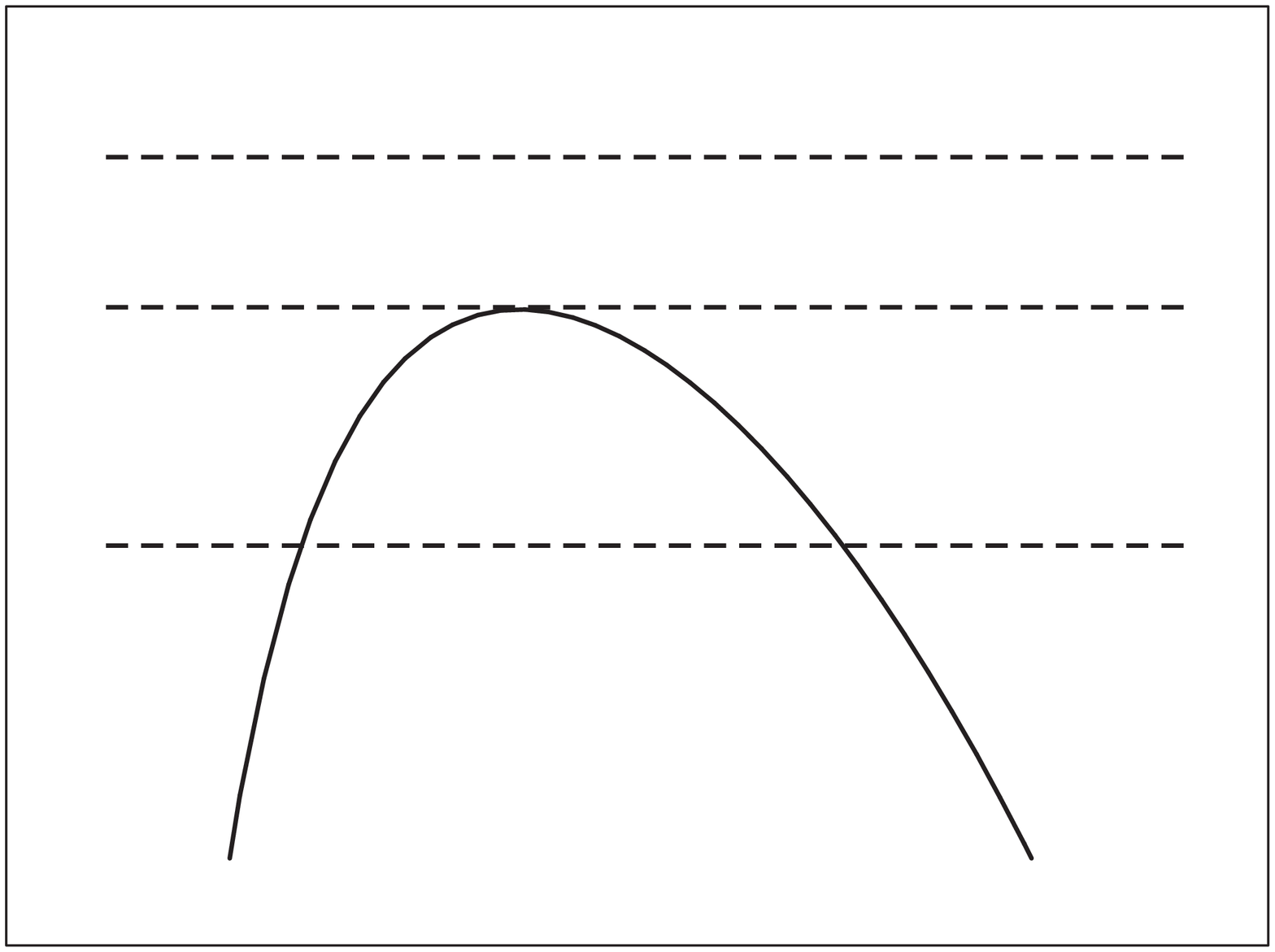}&
\includegraphics[height=3.35cm]{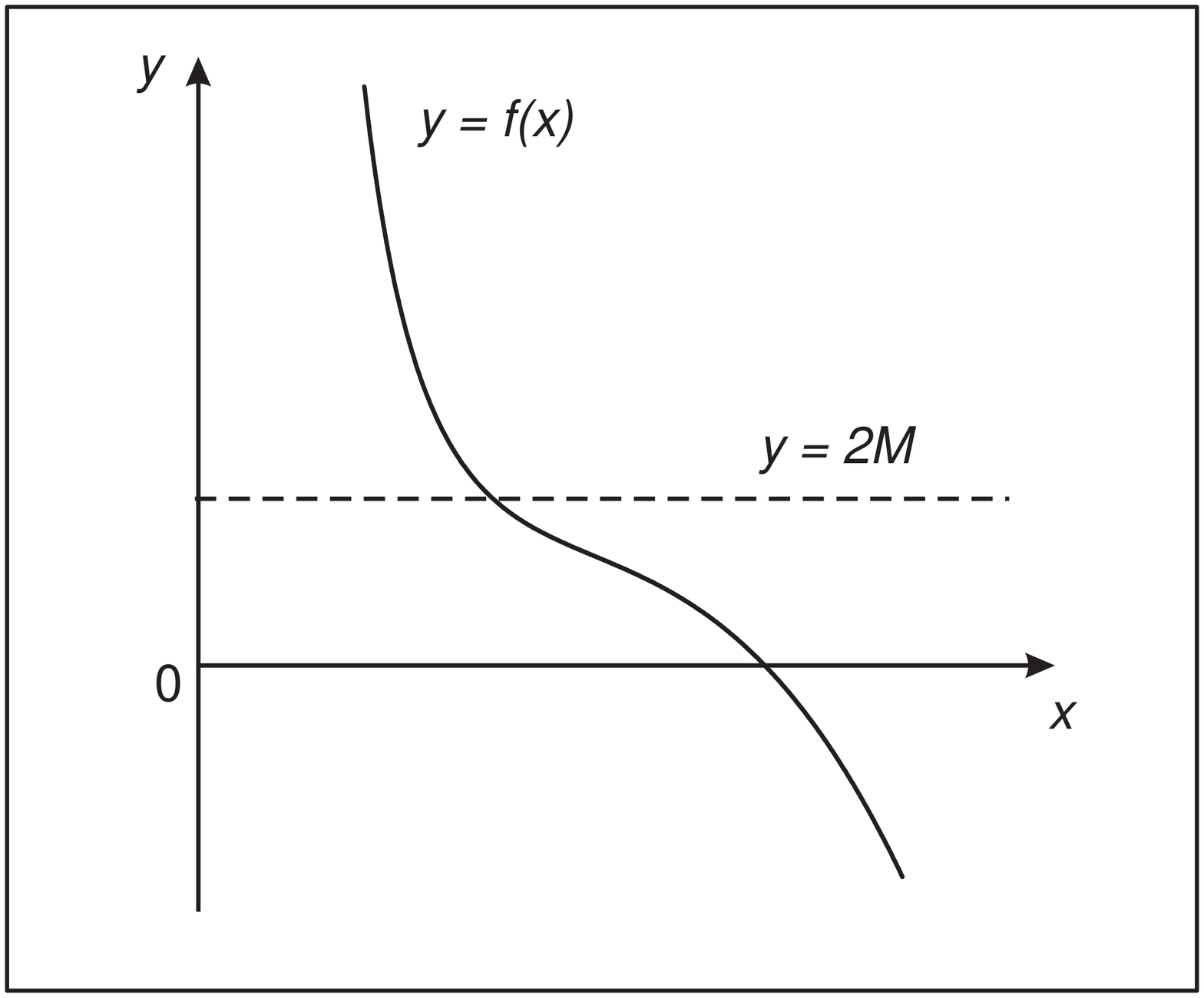}
&\includegraphics[height=3.35cm]{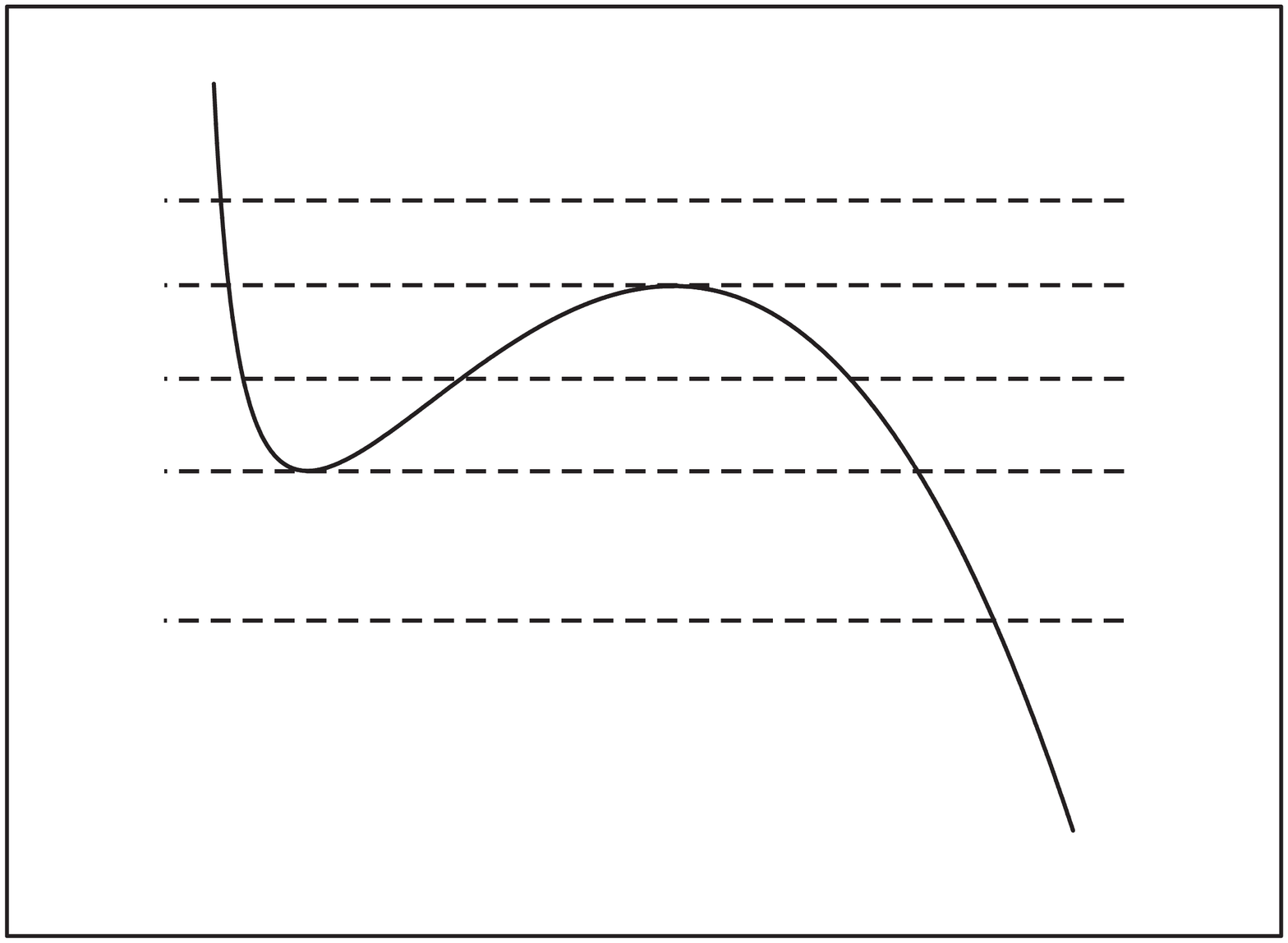}&
\includegraphics[height=3.35cm]{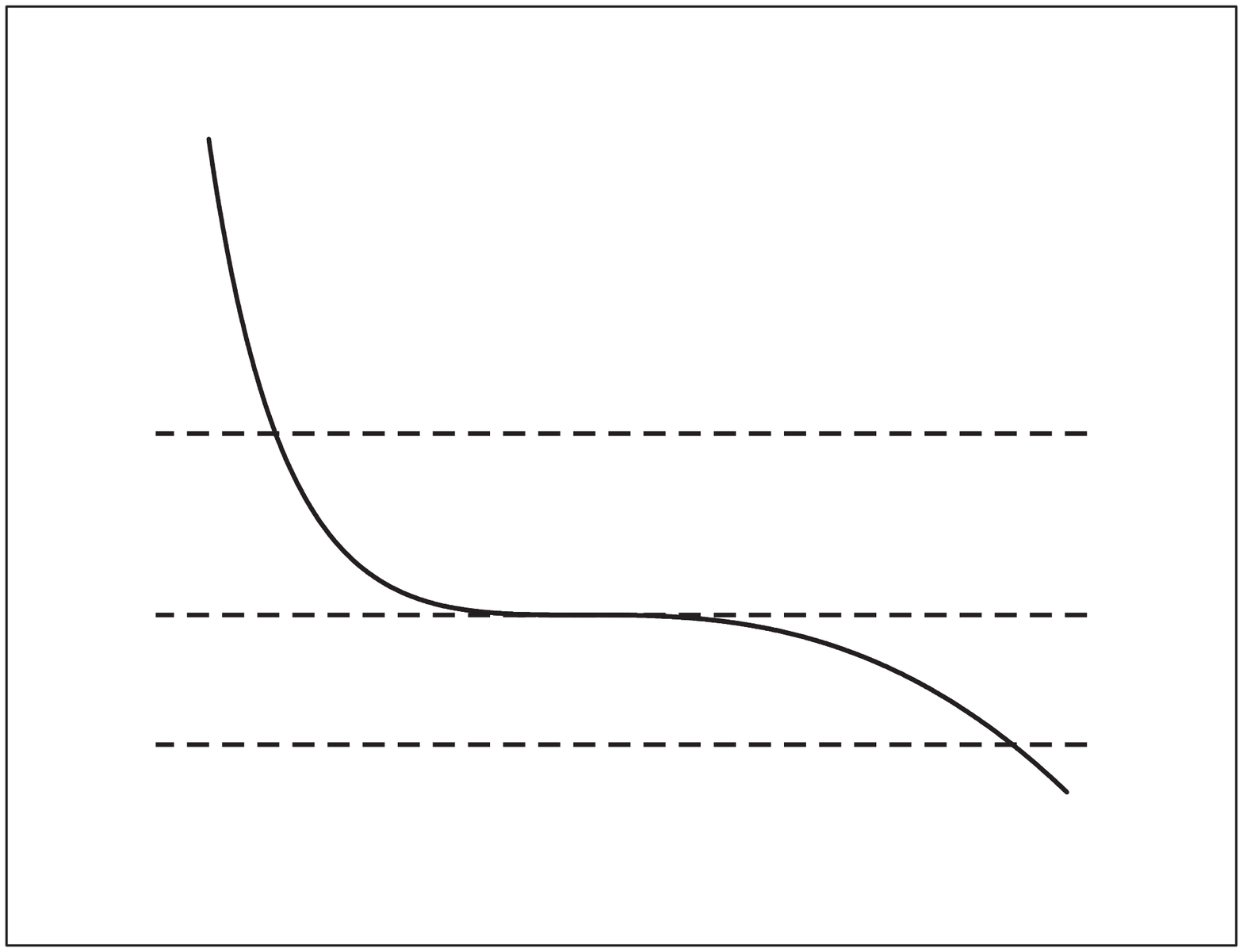}\cr
$(1e)$&$(1f)$&$(1g)$&$(1h)$\cr}
\caption{$\Lambda>0$. Sketches of the auxiliary function
$f(r)=  -\frac{1}{3}\Lambda r^3 + r + {Q_{m}^2} r^{-1} +
{2Q_{m}^2} \bar{q} r^{-3}$ for positive cosmological constant
$\Lambda>0$. At the points $r_{h}$, in which the horizontal mass line
$y=2M$ crosses the curve of the function $y=f(r)$, the
metric function $N(r)$ takes zero values, i.e., the spheres
$r=r_{h}$ are horizons.
Panels $(1a)$, $(1b)$, $(1c)$
$(1d)$, and $(1e)$  show typical cases
for which ${\bar q}<0$, and panels
$(1f)$, $(1g)$, and $(1h)$ illustrate the cases for
$\bar{q}\geq0$.
In panel $(1a)$
the plot of the function $f(r)$
displays two inflexion points and three extrema.
Depending on the mass $M$, and
thus on the mass line
$y=2M$, one can find, counting the intersections from
top to bottom, zero, one, two, three,
four, three, and two cross-points with $y=f(r)$,
respectively. So the spacetime can contain
the following number of horizons: zero, one
double, two simple,
two simple plus one double (corresponding to a maximum),
four simple, one simple plus one double (corresponding to a minimum)
plus one simple, and two simple one of them being a distant one,
horizons. In the
case when the heights
of the two maxima are equal one gets
two double horizons.
Also, it is possible that the left maximum of the
auxiliary function is lower that the right maximum.
The number of horizons remains
the same and we find that
there is no need to display this curve.
Panel $(1b)$ illustrates the
solutions with two inflexion points and one maximum. The
corresponding number of horizons can be zero, one double, and
two simple horizons.
Panel $(1c)$ shows solutions
for which the second inflexion point coincides with one
minimum and one maximum. This case admits zero, one
double, two simple, one simple plus one triple,
and
two simple one of them is a distant one,
horizons.
Panel $(1d)$ corresponds to
the situation, when all inflexion and extremal
points coincide, i.e.,
two maxima
coincide with a minimum and two inflexion points.
This case admits zero, one
quadruple, and
two simple horizons one of them is a distant one.
Panel $(1e)$
illustrates the case, when there are no inflexion points and there is
only one extremum, a maximum. In this case
there are zero, one double, and two simples horizons,
one of them is a distant one.
Panel $(1f)$
depicts the case where one has an
inflexion point and there are no extrema.  In this case
there is only
one simple horizon.
Panel $(1g)$
corresponds to
the situation
where there is one inflexion point and a pair of extrema,
a
minimum comes first, and a maximum comes second.
Depending on the
mass $M$ one can obtain one, one plus a double one,
three simple, a double one plus one distant simple horizon,
and one distant simple horizon.
Panel $(1h)$
illustrates the case, when the minimum, the
inflexion point, and the maximum coincide. Here
the possibilities are:
one simple horizon, a triple horizon, and
one distant simple horizon.
 }\label{figc}
\end{figure*}

The case $\Lambda >0$ is subdivided into
$\bar{q}<0$ and $\bar{q}\geq0$.
Moreover,
when the cosmological constant is positive, $\Lambda >0$,
the discriminant $\Delta$
in Eq.~(\ref{revN04}) can be positive, vanishing, or
negative. This means
that there is a critical value of the cosmological constant
$\Lambda_{c}$ which we have found previously,
see  Eq.~(\ref{lambdac}), given by
$\Lambda_{c} = \frac{{Q_{m}^2}}{972\,{\bar q}^2}$,
with $\Lambda_{c}>0$.
So, in the analysis of the case $\Lambda >0$
we have to distinguish
the cases $0<\Lambda<\Lambda_{c}$,
$\Lambda=\Lambda_{c}$ and $\Lambda>\Lambda_{c}$.
All panels and plots for this $\Lambda >0$
case are shown in
Fig.~\ref{figc}.

\subsection{The case $\bar{q}<0$ and $\Lambda >0$}

Here, Eqs.~(\ref{revN00xx}), (\ref{revN010}), and (\ref{revN01})
yield
\begin{gather}
f(r) =  -\frac{|\Lambda| r^3}{3} + r +
\frac{{Q_{m}^2}}{r} - \frac{{2Q_{m}^2} |\bar{q}|}{r^3}\,,
\label{revN00xxq<0Lambda>0}\\
|\Lambda| r^6 -r^4 + {Q_{m}^2} r^2 - 6 {Q_{m}^2} |\bar{q}| =0\,,
\label{revN010q<0Lambda>0}\\
-|\Lambda| r^6 + {Q_{m}^2} r^2 - 12{Q_{m}^2} |\bar{q}| =0\,,
\label{revN01q<0Lambda>0}
\end{gather}
respectively. The analysis proceeds as in the previous
section.

\vskip 0.3cm
\noindent {\it (1) $0<\Lambda<\Lambda_{c}$}

\noindent In this case, the discriminant given by
Eq.~(\ref{revN04}) is positive, there are
three real different roots $Y_{1}$, $Y_{2}$ and $Y_{3}$, and
two of them are positive (see Eq.~(\ref{revN05})).
These two positive real roots
$r_{i1}$ and $r_{i2}$
of Eq.~(\ref{revN01q<0Lambda>0}), are such that $r_{i1}\neq r_{i2}$,
i.e., there are
two noncoinciding inflexion points of the function $f(r)$.
As for
the roots $X_{1}$, $X_{2}$, $X_{3}$ of
Eq.~(\ref{revN010q<0Lambda>0})
(see Eq.~(\ref{revN051})
for the analysis of their signs)
one finds that there are three possibilities:
there is the possibility of
three different real positive roots, the possibility of
one real positive root, and the possibility of
three real positive roots two of
which coincide.
The case when three real roots coincide does not
give two inflexion points, it is a degenerated case.
As for the function $f(r)$ itself
we find that
$f(0)=-\infty$ and $f(\infty)=-\infty$.
Thus, taking into account altogether
there are three
main situations.

The first situation is described in panel $(1a)$ of Fig.~\ref{figc}.
The function has one local minimum, two local maxima,
and two inflexion points
between the corresponding maxima and minimum. The horizontal
mass line can cross this plot
zero, one, two, three and four times.
This means, that this model can
admit
zero, one
double, two simple,
two simple plus one double (corresponding to a maximum),
and four simple
horizons.
In the
case when the heights
of the two maxima are equal one gets
two double horizons. There is also  a situation where
two simple plus one double (corresponding to a minimum)
horizon exist, and a different situation where there are
two simple horizons, one of them is a distant one.

The second situation is described in panel $(1b)$ of Fig.~\ref{figc},
it has only local maximum. Thus, this model admits zero, one double,
and two simple horizons.

The third situation is described in panel $(1c)$ of Fig.~\ref{figc},
it is characterized by the coincidence of the minimum with one maximum
and one inflexion point.  This case admits zero, one double, two
simple, and two horizons, one is simple and the other is triple. There
is also a different situation where two simple horizons exist, one of
them is a distant one.

Thus, in the case $\bar{q}<0$ and
$0<\Lambda<\Lambda_{c}$ we obtain models with
zero, one, two, three, or four horizons.

\vskip 0.3cm
\noindent {\it (2) $\Lambda=\Lambda_{c}$}

\noindent
When the discriminant given by Eq.~(\ref{revN04}) vanishes,
Eq.~(\ref{revN01q<0Lambda>0}) takes the simple multiplicative form
$(r^2-18
|\bar{q}|)^2 (r^2+36 |\bar{q}|)=0$. Clearly, for negative $\bar{q}$ there
are two coinciding real positive roots of this equation,
$r_i=3\sqrt{2|\bar{q}|}$, and thus, there is one double inflexion
point for the function $f(r)$.  If, in addition,
$|\bar{q}|=Q_m^2/18$, i.e.,
$\bar{q}=-Q_m^2/18$,
Eq.~(\ref{revN010q<0Lambda>0}) converts into $(r^2-Q^2_m)^3=0$, and thus, two
maxima coincide with the minimum at $r=|Q_m|$ and coincides with
double inflexion point $r_i=3\sqrt{2|\bar{q}|}$.
We also have $f(0)=-\infty$ and $f(\infty)=-\infty$.  The sketch of the
corresponding function $f(r)$ is represented in panel $(1d)$ of the
Fig.~\ref{figc}.  It can be considered as a limiting case of panel
$(1b)$.  This case admits zero, one quadruple, and two simple
horizons, one of them is a distant one.

\vskip 0.3cm
\noindent {\it (3) $\Lambda>\Lambda_{c}$}

\noindent  When
the discriminant  given by Eq.~(\ref{revN04}) is negative, there is
one real root $Y_{1}$, and it is negative.
So, there are no real
positive roots and no inflexion points of the function $f(r)$.
Again $f(0)=-\infty$ and $f(\infty)=-\infty$.
There is only one
variant with one extremum, a maximum of the function $f(r)$.
See
panel $(1e)$ of Fig.~\ref{figc}.
In this case
there are zero, one double, and two simples horizons,
one of them is a distant one.

\subsection{The case $\bar{q}\geq0$ and $\Lambda >0$}

For $\bar q =0$, Eqs.~(\ref{revN00xx}),
(\ref{revN010}), and (\ref{revN01})
are given by
\begin{gather}
f(r) =  -\frac{|\Lambda| r^3}{3} + r +
\frac{{Q_{m}^2}}{r} \,,
\label{revN00xxq=0Lambda>0}\\
|\Lambda| r^6 -r^4 + {Q_{m}^2} r^2  =0\,,
\label{revN010q=0Lambda>0}\\
-|\Lambda| r^6 + {Q_{m}^2} r^2  =0\,,
\label{revN01q=0Lambda>0}
\end{gather}
respectively.

One has one inflexion point
$r_{i}=\left(\frac{{Q_{m}^2}}{|\Lambda|} \right)^{\frac{1}{4}}$,
there are at most two extrema at
$r_{{\rm extr1}}=
\sqrt{\frac{1}{2|\Lambda|}\left(1+ \sqrt{1-4{Q_{m}^2} |\Lambda|} \right)}$
and
$r_{{\rm extr2}}=
\sqrt{\frac{1}{2|\Lambda|}\left(1- \sqrt{1-4{Q_{m}^2} |\Lambda|} \right)}$,
and
at $4{Q_{m}^2} |\Lambda|=1$ the extrema coincide with the inflexion point.
Also, $f(0)=+\infty$, $f(\infty)=-\infty$. For $\bar{q}=0$, at $r=0$
one has a horizon that is singular.
Apart from this initial analysis $\bar q=0$ and
$\bar q>0$ have the same type of behavior.

We analyze
it in the following for a generic $\bar q>0$.
Here, Eqs.~(\ref{revN00xx}), (\ref{revN010}), and (\ref{revN01})
yield
\begin{gather}
f(r) =  -\frac{|\Lambda| r^3}{3} + r +
\frac{{Q_{m}^2}}{r} + \frac{{2Q_{m}^2} |\bar{q}|}{r^3}\,,
\label{revN00xxq>0Lambda>0}\\
|\Lambda| r^6 -r^4 + {Q_{m}^2} r^2 + 6 {Q_{m}^2} |\bar{q}| =0\,,
\label{revN010q>0Lambda>0}\\
-|\Lambda| r^6 + {Q_{m}^2} r^2 + 12{Q_{m}^2} |\bar{q}| =0\,,
\label{revN01q>0Lambda>0}
\end{gather}
respectively.

\vskip 0.3cm
\noindent {\it (1)  $0<\Lambda<\Lambda_{c}$}

\noindent
Here the discriminant given in Eq.~(\ref{revN04}) is positive and
there are three real roots $Y_{1}$, $Y_{2}$ and $Y_{3}$.  Since the
product of the roots is positive and the sum is equal to zero, two of
them should be negative and one positive.  This means that there
exists only one positive real value $r_{i}$, the root of
Eq.~(\ref{revN01q>0Lambda>0}), indicating one inflexion point of the function
$f(r)$. The product of the roots $X_{1}$, $X_{2}$, $X_{3}$ is negative
and their sum is positive.  Thus, one has two possibilities: first,
there is a pair of complex conjugated roots and one negative real
root, and second, there are two positive and one negative real
roots. The case with three negative real roots should be
excluded. Taking into account that in this case $f(0)=+\infty$ and
$f(\infty)=-\infty$, we see that there are three possible situations.

The first situation is described in panel $(1f)$ of Fig.~\ref{figc}.
There is one
inflexion point but there are no extrema.
The mass line crosses the curve only once, and this
means that inevitably there is one and only one simple horizon.

The second situation is described in panel $(1g)$ of Fig.~\ref{figc}.
The curve $f(r)$ has one
local minimum, one local maximum  and one inflexion point between
them.
Depending on the
mass $M$ one can obtain one, one plus a double one,
three simple, a double one plus one distant simple horizon,
and one distant simple horizon.

The third
situation is described in panel $(1h)$ of Fig.~\ref{figc}.
This case is degenerated, i.e., the
maximum, the minimum and the inflexion points coincide.
One can have
one simple horizon, a triple horizon, and
one distant simple horizon.

Thus, in the case $\bar q>0$ and
$0<\Lambda<\Lambda_{c}$  we obtain  models
with at least one horizon, and two or three horizons can appear
for specific values of the mass $M$.

\vskip 0.3cm
\noindent {\it (2) $\Lambda=\Lambda_{c}$}

\noindent The discriminant in Eq.~(\ref{revN04}) vanishes.
There are
three real roots: one positive and two coinciding negative roots.
Thus, there is one real positive root $r_{1i}$, and so one inflexion
point of the function $f(r)$. Since $f(0)=+\infty$ and
$f(\infty)=-\infty$, the plots of this function $f(r)$ are
given in panels $(1f)$, $(1g)$ and
$(1h)$ of Fig.~\ref{figc}.

\vskip 0.3cm
\noindent {\it (3) $\Lambda>\Lambda_{c}$}

\noindent The discriminant in Eq.~(\ref{revN04}) is negative.
There is
one real positive root, say, $Y_{1}$, so one positive real root
$r_{1i}$, and thus
one inflexion point of the function $f(r)$. Since
$f(0)=+\infty$ and $f(\infty)=-\infty$, we deal with one of the
situations described in panels
$(1f)$, $(1g)$, and $(1h)$ of Fig.~\ref{figc}.

\section{Exact solutions with zero
cosmological constant, $\Lambda =0$}
\label{secxclass}

\begin{figure*}[t]
\halign{\hfil#\hfil&\;\hfil#\hfil&\;\hfil#\hfil&\;\hfil#\hfil\cr
\includegraphics[height=4cm]{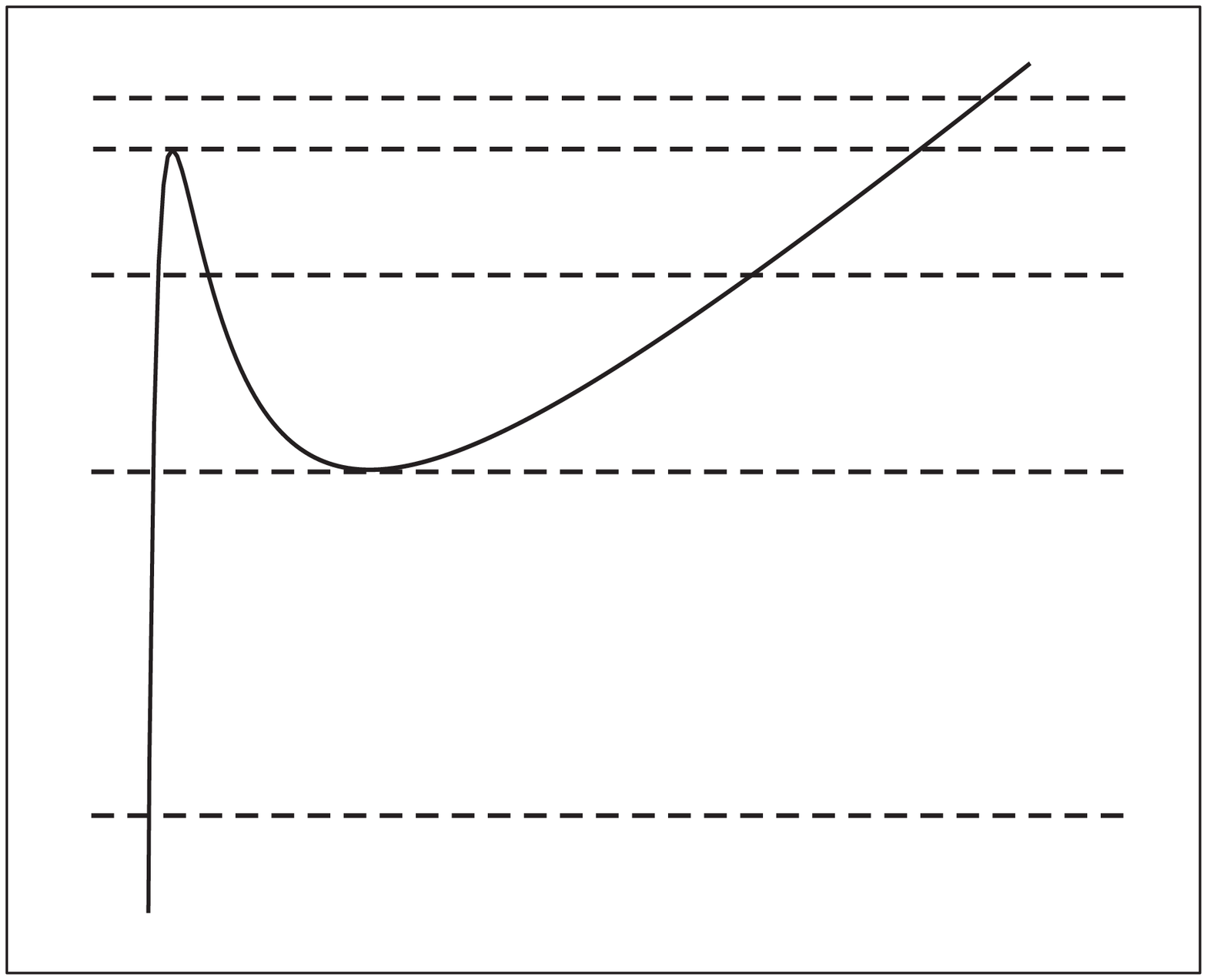}&
\includegraphics[height=4cm]{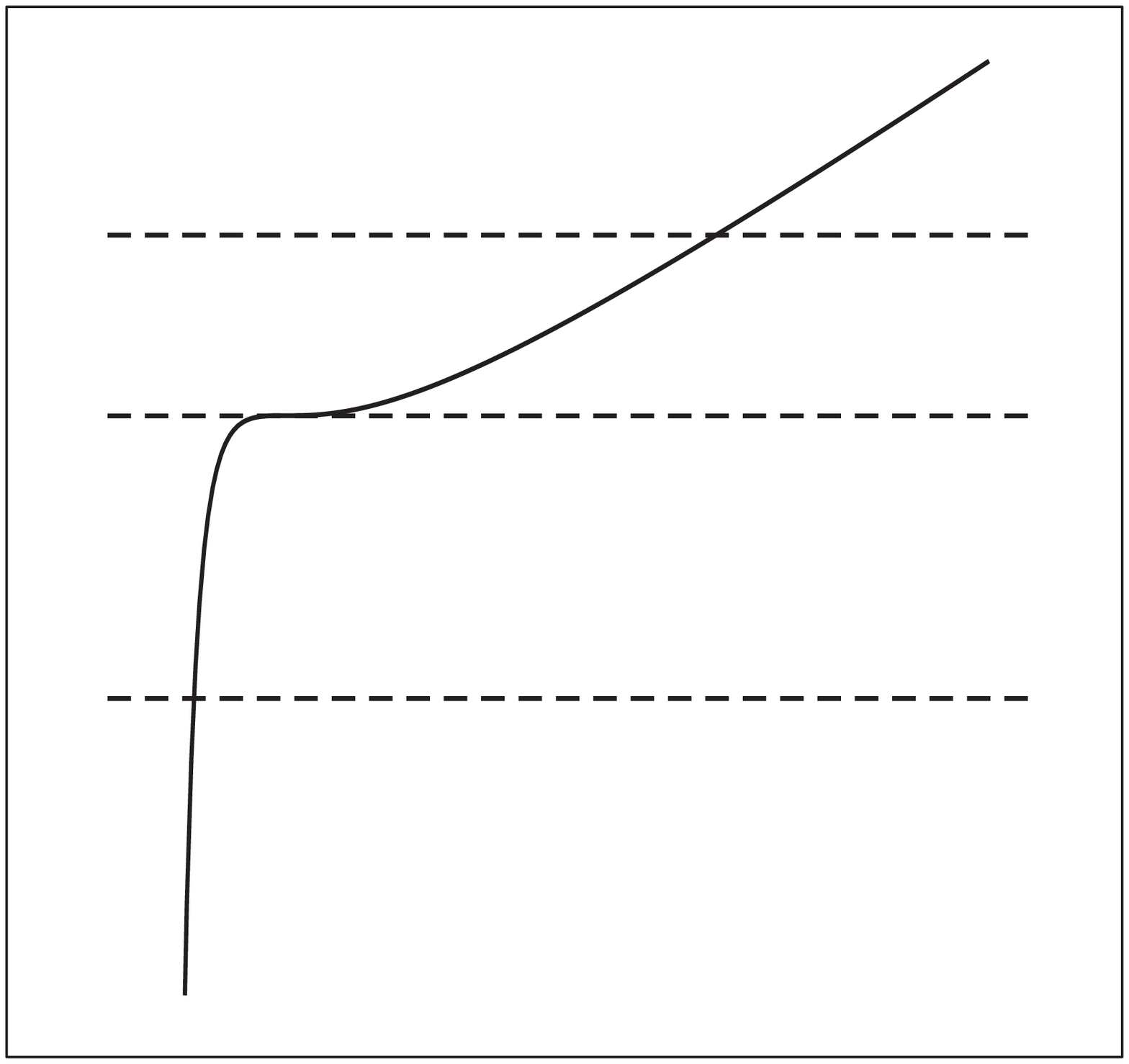}&
\includegraphics[height=4cm]{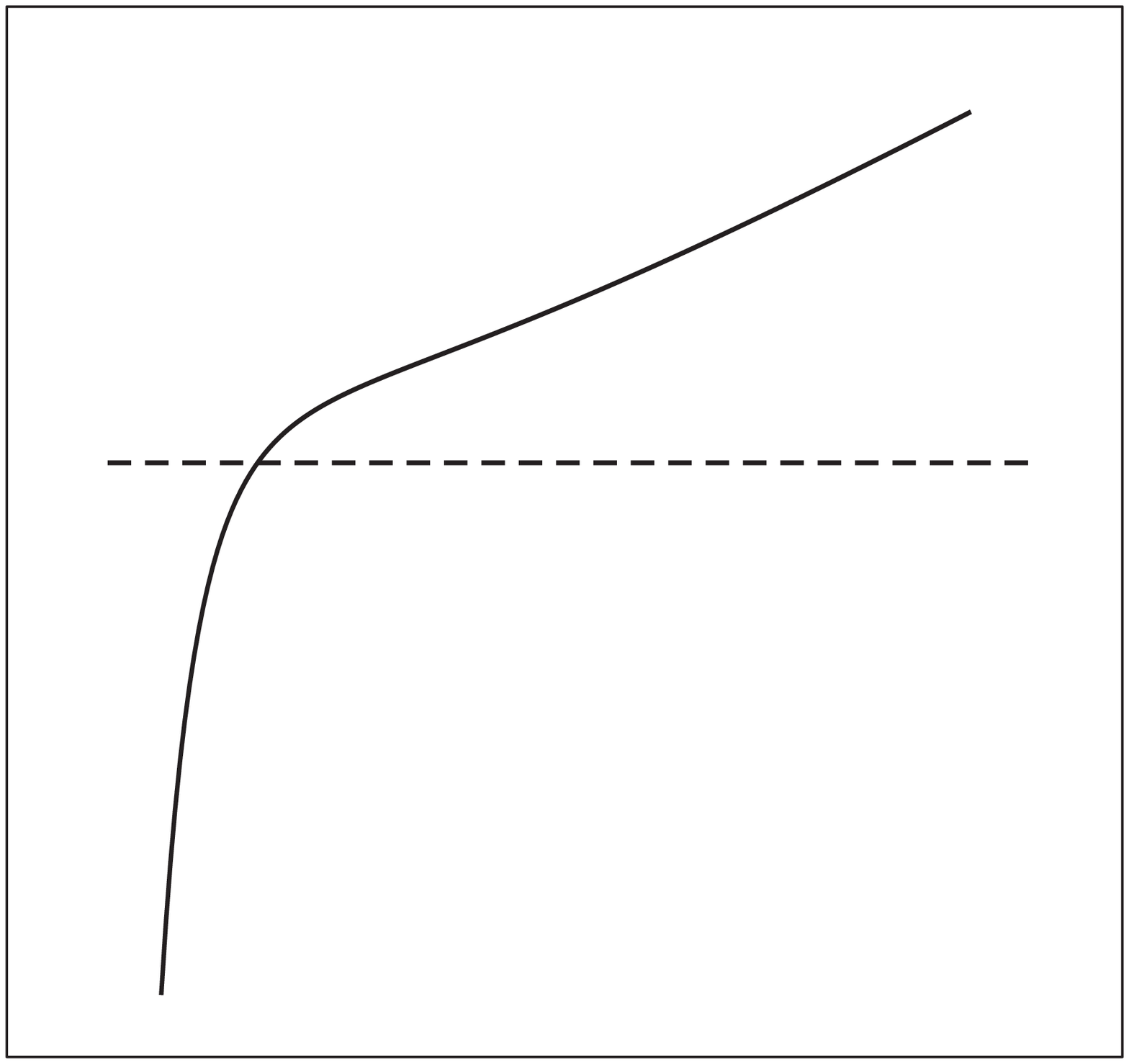}&
\includegraphics[height=4cm]{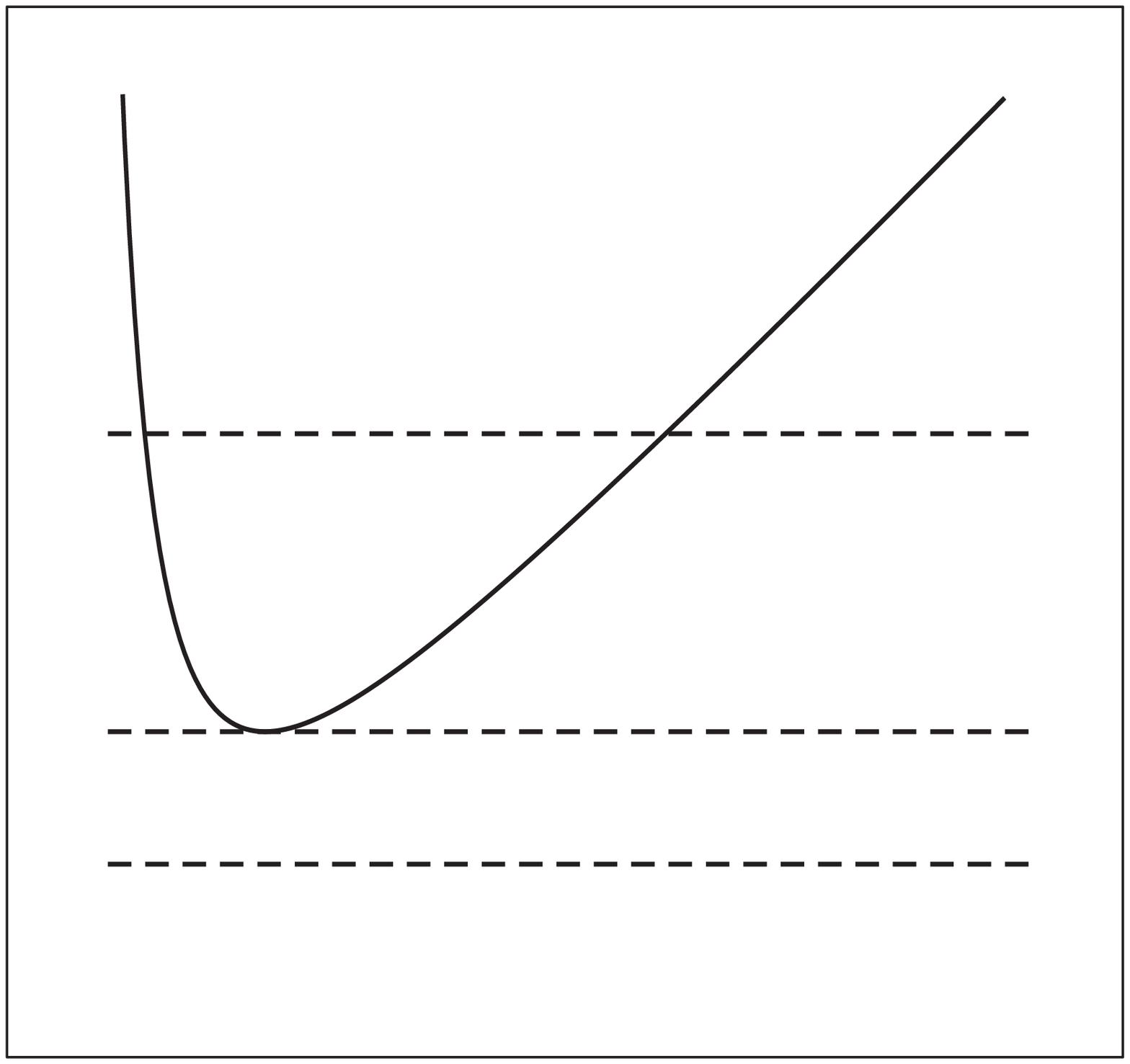}\cr
$(2a)$&$(2b)$&$(2c)$&$(2d)$\cr} \caption{
$\Lambda=0$. Sketches of the auxiliary function $f(r)= r + Q_{m}^2
r^{-1} + {2Q_{m}^2} \bar{q} r^{-3}$ for zero cosmological constant
$\Lambda=0$. At the points $r_{h}$, in which the horizontal mass line
$y=2M$ crosses the curve of the function $y=f(r)$, the metric function
$N(r)$ takes zero values, i.e., the spheres $r=r_{h}$ are horizons.
The plots on panels $(2a)$, $(2b)$, $(2c)$ show typical cases for
which ${\bar q}<0$, and panel $(2d)$ illustrates the cases for
$\bar{q}\geq0$.  In panel $(2a)$ the plot of the function $f(r)$
displays one inflexion point and two extrema.  Depending on the mass
$M$, and thus on the mass line $y=2M$, one can find one, one double
plus one simple, three simple, one simple plus one double, and one
simple horizon.  In the case the extrema coincide one has a quadruple
horizon, not shown in panel.  In panel $(2b)$ the inflexion point
coincides with an extremum.  One can find one simple horizon. In panel
$(2c)$ there is one inflexion point and no extrema.  One can find one
simple horizon.
In panel $(2d)$
there are no inflexion points and there is one extremum, a minimum. In
this case there are zero, one double, and two simple horizons.
} \label{figb}
\end{figure*}

The case $\Lambda =0$ is subdivided into
$\bar{q}<0$ and $\bar{q}\geq0$.
All panels and plots for this $\Lambda =0$
case are shown in
Fig.~\ref{figb}.

\subsection{The case $\bar{q}< 0$ and $\Lambda = 0$}

Here, Eqs.~(\ref{revN00xx}), (\ref{revN010}), and (\ref{revN01})
yield
\begin{gather}
f(r)= r + \frac{{Q_{m}^2}}{r} -
\frac{{2Q_{m}^2}}{r^3}|\bar{q}| \,,
\label{revN2}\\
r^4 - {Q_{m}^2} r^2 + 6{Q_{m}^2}
|\bar{q}| = 0 \,,
\label{revN211}\\
r^2 = 12 |\bar{q}|
\label{revN222}\,,
\end{gather}
respectively.
Clearly, the case $\Lambda=0$ is much simpler for classification than
$\Lambda > 0$, since Eq.~(\ref{revN222})
is now a  quadratic equation  (instead of bicubic)
for obtaining
the inflexion point, and Eq.~(\ref{revN211})
is a biquadratic equation
for obtaining of extrema, instead of the corresponding bicubic
equation.  The auxiliary function $f(r)$,
Eq.~(\ref{revN2}), has the following asymptotic
properties: $f(0){=}{-}\infty$, $f(\infty){=}\infty$.
There is one
inflexion point at $r {=} \sqrt{12 |\bar{q}|}$.  The parameter
$\frac{24\bar{q}}{{Q_{m}^2}}$ is now the critical parameter of the model,
and we have three intrinsic situations.
\noindent

The first situation
is displayed on panel $(2a)$ of Fig.~\ref{figb}
and corresponds to the case when $\frac{24\bar{q}}{{Q_{m}^2}}<1$.
There are two extrema: the maximum at $r_{\rm extr1}
{=}\sqrt{\frac{{Q_{m}^2}}{2}-
\sqrt{\frac{{Q_{m}^4}}{4}{-}6{Q_{m}^2}
|\bar{q}|}}$, and minimum at $r_{\rm extr2}{=}
\sqrt{\frac{{Q_{m}^2}}{2} +
\sqrt{\frac{{Q_{m}^4}}{4}{-}6{Q_{m}^2} |\bar{q}|}}$).
There are a first
and a second distinguishing masses $M_1 = \frac12 f(r_1)$ and $M_2 =
\frac12 f(r_2)$. When $M<M_2$ there is one simple horizon. When
$M=M_2$, there are one simple horizon and one double distant
horizon. When $M_2<M<M_1$ there are three simple horizons. When
$M=M_1$ there are one double horizon and one simple distant
horizon. When $M>M_1$ there is one simple distant horizon.

The second situation
is displayed on panel $(2b)$ of Fig.~\ref{figb}
and corresponds to the case when $\frac{24\bar{q}}{{Q_{m}^2}}=1$.
Now the maximum, minimum and inflexion points coincide. We have now only
one distinguishing mass $M_{\rm t}$. When
$M<M_{\rm t}$, there is one simple horizon. When $M=M_{\rm t}$ the
horizon is triple. When $M>M_{\rm t}$, there is one simple distant
horizon.

The third situation
is displayed on panel $(2c)$ of Fig.~\ref{figb}
and corresponds to the case when $\frac{24\bar{q}}{{Q_{m}^2}}>1$.
Here there are no extrema.
There are no distinguishing masses,
and there is
one simple horizon.

\subsection{The case  $\bar{q}\geq0$ and $\Lambda = 0$}

The case $\bar{q}=0$ is simple.
Eqs.~(\ref{revN00xx}), (\ref{revN010}), and (\ref{revN01})
yield here
\begin{gather}
f(r)= r + \frac{{Q_{m}^2}}{r}  \,,
\label{revN210}\\
-r^4 + {Q_{m}^2} r^2 = 0  \,,
\label{revN211qq0}\\
r^2 =0
\label{revN222qq0}\,,
\end{gather}
respectively.
So $f(r)$
has no inflexion points
and possesses one minimum at $r_{\rm
min}{=}Q_{m}$.
There is one distinguishing mass, $M_1=\frac12 f(r_{\rm
min})=Q_{m}$. When $M<M_1$, there are no
horizons. When $M=M_1$, we obtain one double horizon. When $M>M_1$,
there are two simple horizons, see the sketch on panel $(2d)$
of Fig.~\ref{figb}.

For the case $\bar{q}> 0$ and $\Lambda = 0$,
Eqs.~(\ref{revN00xx}), (\ref{revN010}), and (\ref{revN01})
yield
\begin{equation}
f(r)= r + \frac{{Q_{m}^2}}{r} +
\frac{{2Q_{m}^2} \bar{q}}{r^3} \,,
\label{revN21}
\end{equation}
\begin{equation}
-r^4 + {Q_{m}^2} r^2 + 6{Q_{m}^2}
\bar{q} = 0  \,,
\label{revN211qq}
\end{equation}
\begin{equation}
r^2 + 12 \bar{q}=0
\label{revN222qq}\,,
\end{equation}
respectively.
Clearly, Eq.~(\ref{revN222qq}) has no real roots
and thus there are no
inflexion points. Eq.~(\ref{revN211qq})
has one real positive root
$r_{\rm min}{=}\sqrt{\frac{{Q_{m}^2}}{2} {+}
\sqrt{\frac{{Q_{m}^4}}{4}{+}6{Q_{m}^2} \bar{q}}}$, which gives
the minimum of $f(r)$. Eq.~(\ref{revN21})
has asymptotic values
$f(0){=}\infty$ and  $f(\infty){=}\infty$.
See the sketch on panel $(2d)$
of Fig.~\ref{figb}.
As in the previous case $\bar{q}{=}0$ we can find zero, one
double or two simples horizons.

\section{Exact solutions
with a negative cosmological constant, $\Lambda < 0$}
\label{sec4class}

\begin{figure*}[t]
\halign{\hfil#\hfil&\;\hfil#\hfil&\;\hfil#\hfil&\;\hfil#\hfil\cr
\includegraphics[height=3.6cm]{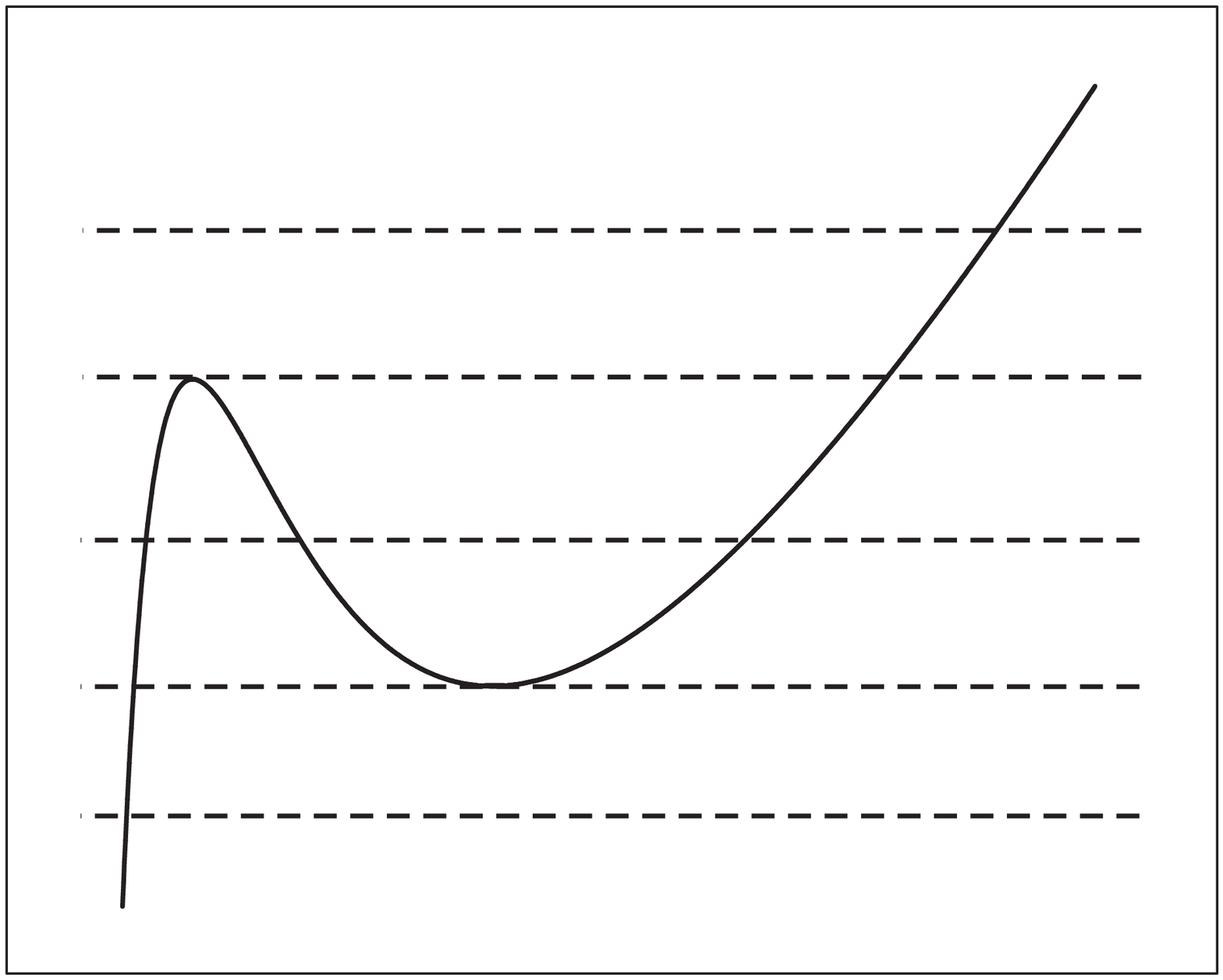}&
\includegraphics[height=3.6cm]{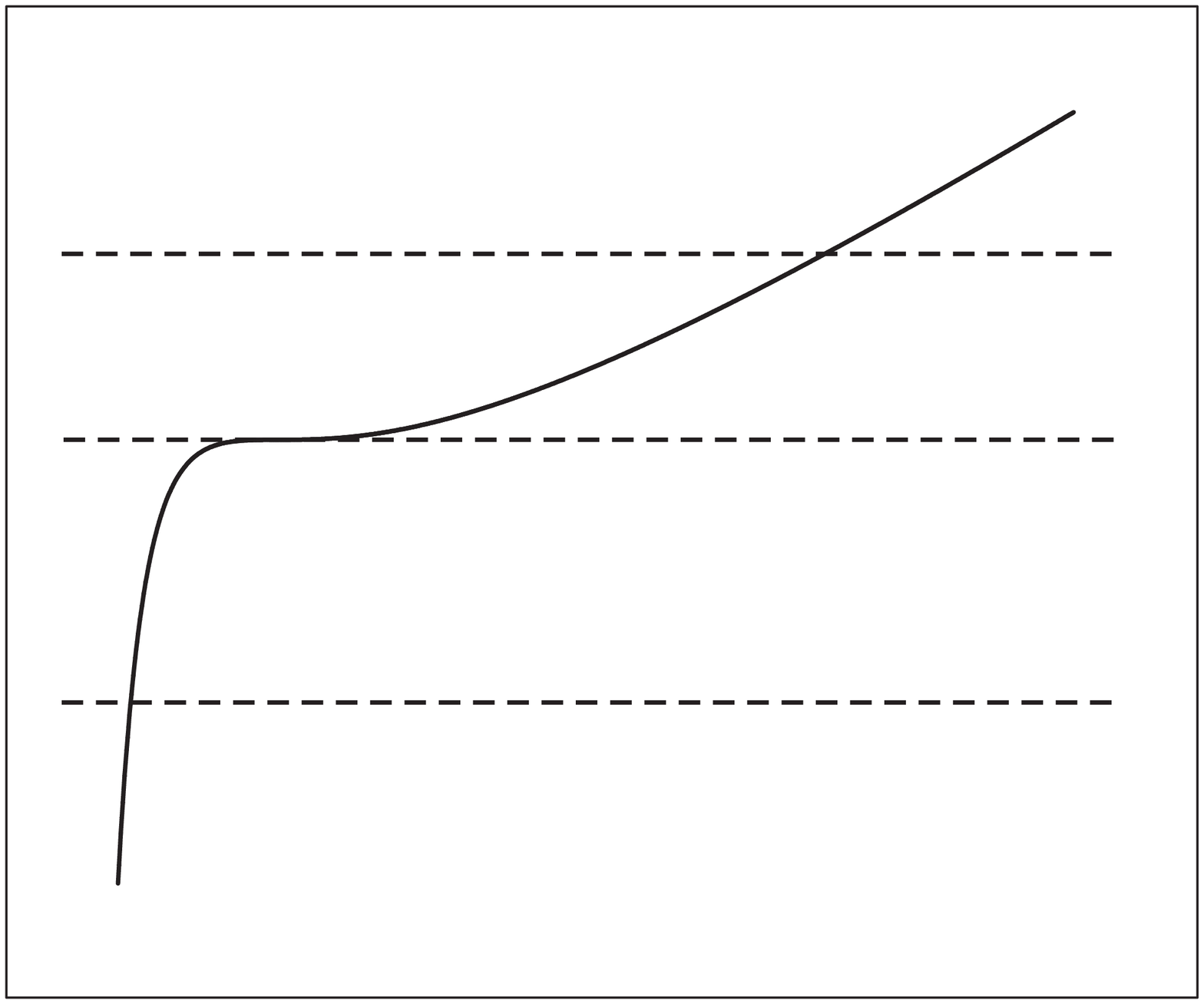}&
\includegraphics[height=3.6cm]{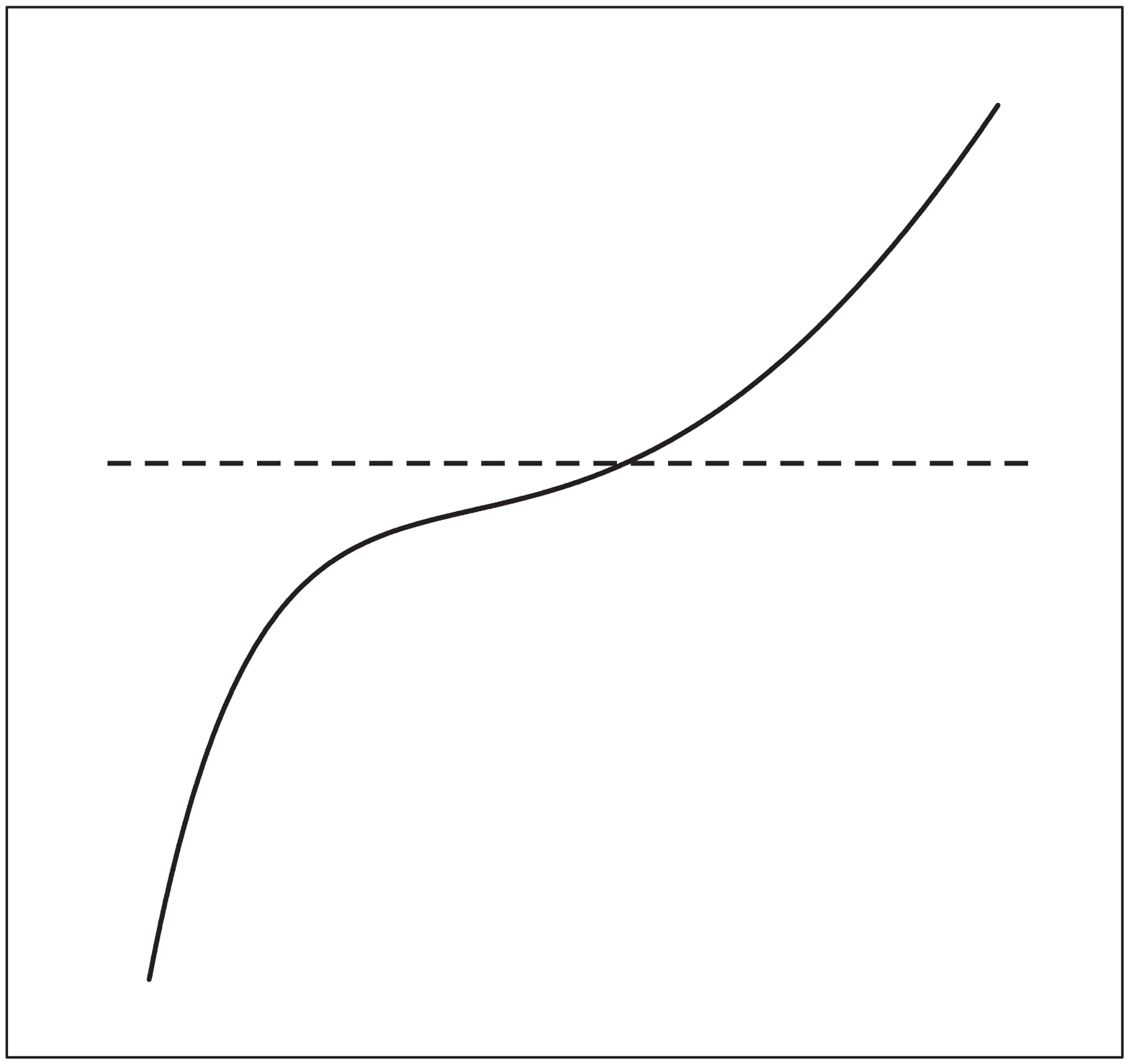}&
\includegraphics[height=3.6cm]{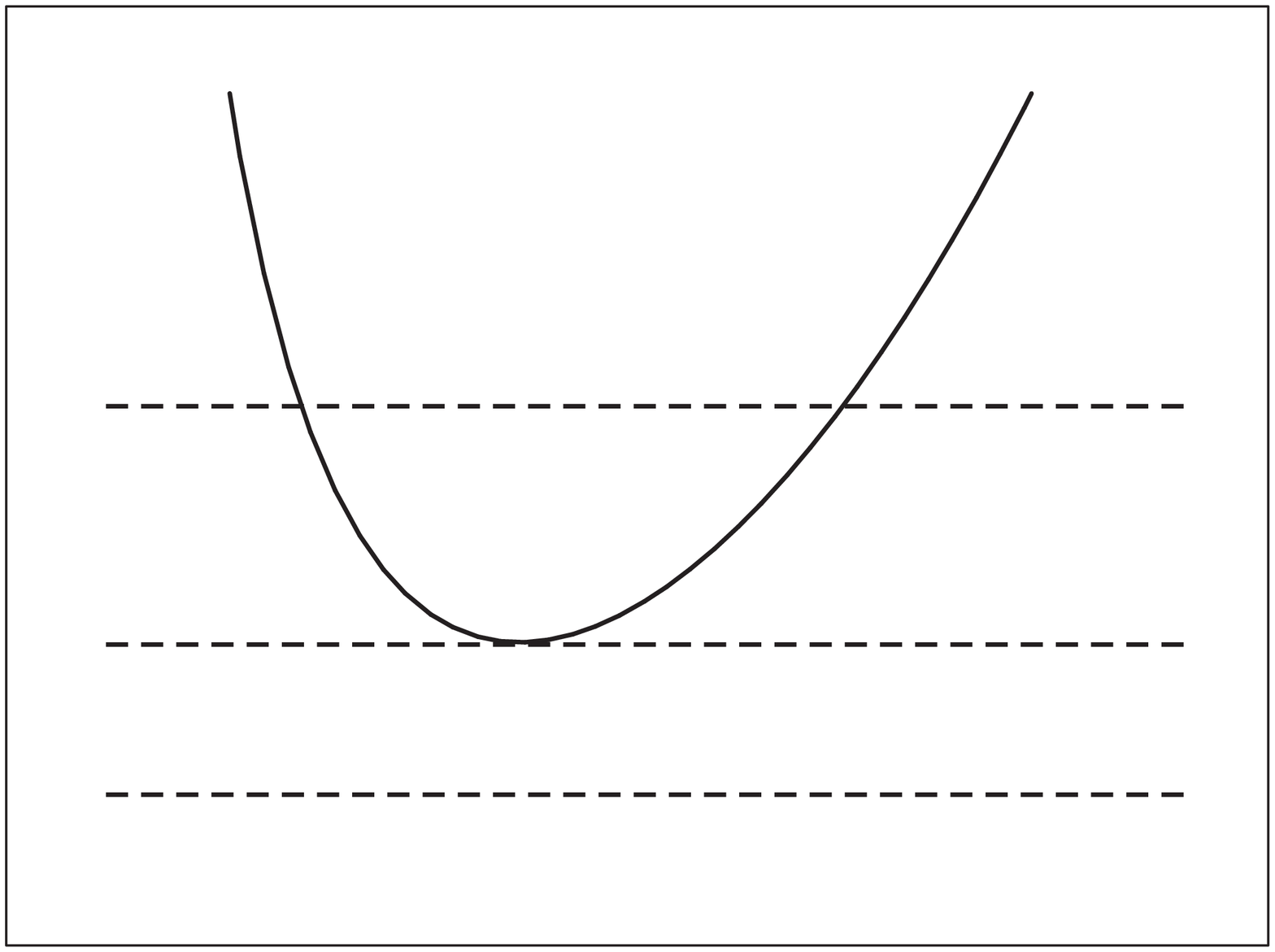}\cr
$(3a)$&$(3b)$&$(3c)$&$(3d)$\cr}
\caption{$\Lambda<0$.
Sketches of the auxiliary function $f(r){=} {-}\frac{1}{3}\Lambda r^3
{+} r {+} Q_{m}^2 r^{-1} {+} {2Q_{m}^2} \bar{q} r^{-3}$ for
negative cosmological constant $\Lambda<0$. At the points $r_{h}$, in
which the horizontal mass line $y=2M$ crosses the curve of the
function $y=f(r)$, the metric function $N(r)$ takes zero values, i.e.,
the spheres $r=r_{h}$ are horizons.  The plots on panels $(3a)$,
$(3b)$, $(3c)$ show typical cases for which ${\bar q}<0$, and panel
$(3d)$ illustrates the cases for $\bar{q}\geq0$.  In panel $(3a)$ the
plot of the function $f(r)$ displays one inflexion point and two
extrema.  Depending on the mass $M$, and thus on the mass line $y=2M$,
one can find one, one double plus one simple, three simple, one simple
plus one double, and one simple horizon.  In the case the extrema
coincide one has a quadruple horizon, not shown in panel.
In panel $(3b)$ the inflexion point coincides
with an extremum.  One can find one simple horizon.
In the
panel $(3c)$ there is one inflexion point and no extrema.  One can find
one simple horizon.
In panel $(3d)$
there are no inflexion points and there is one extremum, a minimum. In
this case there are zero, one double, and two simple horizons.
} \label{figa}
\end{figure*}

The case $\Lambda <0$ is subdivided into
$\bar{q}<0$ and $\bar{q}\geq0$.
All panels and plots for this $\Lambda <0$
case are shown in
Fig.~\ref{figa}.

\subsection{The case $\bar{q}<0$ and $\Lambda<0$}

Here, Eqs.~(\ref{revN00xx}), (\ref{revN010}), and (\ref{revN01})
yield
\begin{equation}
f(r) =  \frac{|\Lambda| r^3}{3} + r +
\frac{{Q_{m}^2}}{r} - \frac{{2Q_{m}^2} |\bar{q}|}{r^3}\,,
\label{revN00xxq<0Lambda<0}
\end{equation}
\begin{equation}
-|\Lambda| r^6 -r^4 + {Q_{m}^2} r^2 - 6 {Q_{m}^2} |\bar{q}| =0\,,
\label{revN010q<0Lambda<0}
\end{equation}
\begin{equation}
|\Lambda| r^6 + {Q_{m}^2} r^2 - 12{Q_{m}^2} |\bar{q}| =0\,,
\label{revN01q<0Lambda<0}
\end{equation}

\noindent In this case the discriminant $\Delta$ in Eq.~(\ref{revN04})
is negative, thus, there is only one real root, say, $Y_{1}$,
the other two roots $Y_{2}$ and $Y_{3}$ are conjugated complex
numbers. Since the product $Y_{1} Y_{2} Y_{3}$ according to
Eq.~(\ref{revN05}) is positive, the real root $Y_{1}$ should be
positive. This means that there exists one real positive root
$r_{i}{=}\sqrt{Y_{1}}$ and thus the function $f(r)$ has one
inflexion point. Since $f(0)=-\infty$, $f(\infty)=+\infty$, the
plots of the function $f(r)$ can be of the type $(3a)$, $(3b)$
or $(3c)$ of  Fig.~\ref{figa}. All these three
sketches have one inflexion point, and the difference between them
is predetermined by the number of extrema, i.e., by the number of
real positive roots of Eq.~(\ref{revN010q<0Lambda<0}),
$X_{1}$, $X_{2}$, $X_{3}$. Since the sum and the product of
these quantities are negative (see Eq.~(\ref{revN051})), there are
three possibilities: only one real negative root, three negative
real roots, and two positive real roots and one negative real root.

When the positive roots $X_{1}$ and $X_{2}$ do not coincide,
one can find one maximum and one minimum, as on the sketch
depicted on panel $(3a)$ of Fig.~\ref{figa}.
When $X_{1}{=}X_{2}$, three points coincide: the maximum, minimum and
inflexion point, see the sketch displayed on panel $(3b)$ of
Fig.~\ref{figa}.
In the first and second cases there are no
extrema, and we obtain the sketch depicted on panel $(3c)$
of Fig.~\ref{figa}.

According to the sketches
$(3a)$, $(3b)$, $(3c)$ the horizontal mass line can cross the
plot of the function $f(r)$ once, two or three times. This means
that this submodel can admit one simple horizon, three simple
horizons, one triple horizon and two horizons, one of them being
a double horizon. In other words, it is guarantied that there is at least
one horizon in this model for any value of the mass $M$.

\subsection{The case $\bar{q}\geq0$ and
$\Lambda <0$}

In the case $\bar{q}=0$
Eqs.~(\ref{revN00xx}), (\ref{revN010}), and (\ref{revN01})
yield
\begin{equation}
f(r)= \frac{|\Lambda| r^3}{3} + r +
\frac{{Q_{m}^2}}{r} \,,
\label{revN2m}
\end{equation}
\begin{equation}
r^2 \left(-|\Lambda|
r^4-r^2+{Q_{m}^2}\right) =0 \,,
\label{revN211qq<0}
\end{equation}
\begin{equation}
r^2 \left(|\Lambda| r^4+{Q_{m}^2}\right) =0
\label{revN222qq<0}\,,
\end{equation}
respectively.
Clearly, when $\Lambda<0$, there are no inflexion points.
There is
only one extremum, a minimum
given by $r_{{\rm min}}{=}
\sqrt{\frac{1}{2|\Lambda|}\left(\sqrt{1{+}4{Q_{m}^2} |\Lambda|} {-}1
\right)}$. Also
$f(0){=}{+}\infty$ and $f(\infty){=}{+}\infty$.
The plot of the
function $f(r)$ has the form of the sketch displayed on panel
$(3d)$ of Fig.~\ref{figa}. In this case one can
find explicitly the critical value of the mass,
$M_{{\rm c}}$. Using the equality
$M_{{\rm c}}= \frac12 f(r_{{\rm min}})$ one finds
\begin{equation}
M_{{\rm c}}=  \frac13
\left(r_{{\rm min}}+ \frac{{2Q_{m}^2}}{r_{{\rm min}}} \right) \,.
\label{Mc}
\end{equation}
When $M<M_{{\rm c}}$, the mass line does not cross the plot of
$f(r)$, i.e., the corresponding object does not have
horizons.
When the mass exceeds the critical one, $M>M_{{\rm c}}$, the
mass line crosses the plot twice, and two horizons appear.
Finally, when $M{=}M_{{\rm c}}$, there is one double horizon.

In the case $\bar{q}>0$ and $\Lambda<0$,
Eqs.~(\ref{revN00xx}), (\ref{revN010}), and (\ref{revN01})
yield
\begin{equation}
f(r) =  \frac{|\Lambda| r^3}{3} + r +
\frac{{Q_{m}^2}}{r} + \frac{{2Q_{m}^2} |\bar{q}|}{r^3}\,,
\label{revN00xxq>0Lambda<0}
\end{equation}
\begin{equation}
-|\Lambda| r^6 -r^4 + {Q_{m}^2} r^2 + 6 {Q_{m}^2} |\bar{q}| =0\,,
\label{revN010q>0Lambda<0}
\end{equation}
\begin{equation}
|\Lambda| r^6 + {Q_{m}^2} r^2 + 12{Q_{m}^2} |\bar{q}| =0\,,
\label{revN01q>0Lambda<0}
\end{equation}
In this case
the discriminant in Eq.~(\ref{revN04}) is negative
and there is
only one real root. Since the product of the roots is now negative,
the real root is also negative. Thus, there are no real roots,
and so
there are no inflexion points.
There exists a minimum of the function $f(r)$, since
$f(0)=+\infty$ and $f(\infty)=+\infty$. The plot of this function
$f(r)$ is depicted in panel $(3d)$ of Fig.~\ref{figa}.
This model admits two simple or one double horizons. The
horizons can disappear when the mass $M$ is less than the critical
mass, related to the minimal value of the function $f(r)$.

\begin{widetext}

\section{Short resum\'e}
\label{shortresume}

In Table~I we present a summary
of the results of the classification of
the five-parameter family of exact solutions with respect to
number of horizons and their type, i.e.,
whether they are single, double, triple, or
quadruple. We use in this table the following symbols:  we write 0 when
there are no horizons; 1s, 2s,
3s, and 4s means one, two, three, and four simple horizons, respectively;
1d and 2d to denote one and two double horizons, respectively;
the symbol 1t to mean one triple horizon; and 1q to mean one quadruple
horizon. We also use the following
expressions: 1s+1d means that there are one simple and one
double horizon; 2d+2d means two double horizons. The entries in the table
are given by the nonminimal parameter
$\bar{q}$ (for the cases $\bar{q} <0$
and $\bar{q} \geq0$)
and the cosmological constant $\Lambda$
(for the cases $ \Lambda < 0$,  $ \Lambda = 0$,
$0< \Lambda < \Lambda_{c}$,
$\Lambda = \Lambda_{c}$, and $\Lambda > \Lambda_{c}$), where
$\Lambda_{c}$ is the critical value of the cosmological
constant $\Lambda_{c} \equiv
\frac{{Q_{m}^2}}{972\bar{q}^2}$, see Eq.~(\ref{lambdac}).

\begin{table*}[h]
\begin{center}
\begin{tabular}{|c|c|c|c|c|c|c|c||}
  \hline
  \,  & \, $ \Lambda < 0$ & $\Lambda = 0$ & $0< \Lambda < \Lambda_{c}$ \,
& $\Lambda = \Lambda_{c}
  $\, & $\Lambda > \Lambda_{c}$ \\
  \hline
  \,$\bar{q} <0\,$ & 1s; 3s; 1t; 1s+1d &1s; 3s; 1t; 1s+1d &0;
  1d; 2s; 2s+1d; 4s; 2d;
1s+1t; 1q & 0; 1d; 2s; 1s+1t; 1q & 0; 2s; 1d \\
  \hline
 \, $\bar{q}\geq 0$\, & 0; 2s; 1d &0; 2s; 1d &1s; 1s+1d; 3s; 1t & 1s;
1s+1d; 3s; 1t & 1s; 1s+1d;
  3s;  1t \\
\hline

\end{tabular}
\caption{The number of horizons in the five-parameter family of
exact solutions as a function of
the nonminimal parameter $\bar{q}$
and the cosmological constant $\Lambda$.}
\end{center}
\label{table1}
\end{table*}

We have now completed the classification
of the black holes and horizons existent in these
spherical symmetric nonminimal models.
We will give below two examples
where this
classification fits.
These examples are typical cases
for $ \bar q<0$ and $\bar q\geq0$.
They are the Drummond-Hathrell model
and the regular black hole, respectively.
\vskip 0.3cm

\end{widetext}

\section{The Drummond-Hathrell model:
Example of a
nonminimal theory with $\bar q<0$
}\label{sec5}

In \cite{drumhat}, Drummond and Hathrell have
investigated a model in which the parameters of the nonminimal
coupling were calculated in the framework of the one-loop
corrections to QED. These parameters are of the form $q_1=-q$,
$q_2=\frac{13}{5}q$, $q_3=-\frac25 q$, where
$q=\frac{\alpha\lambda^2_{e}}{36\pi}$, $\lambda_{e}$ being
the Compton radius of electron. Clearly, for these
Drummond-Hathrell parameters one obtains that $10q_1+4q_2+q_3=0$,
so it obeys Eq.~(\ref{qsrel}) that we have assume from the start.
Moreover $\bar{q}=-\frac25 q$, so $\bar{q}<0$
and it falls in the case  $\bar{q}<0$ that we have treated
before. Since  $\bar q$ is given
for a given $q$, the two nonminimal
coupling constants, $q$ and $\bar q$, reduce to one
independent coupling constant, $q$, say.
We deal with a four parameter model.

Thus, the Drummond-Hathrell model has
\begin{equation}
\sigma(r) =1\,,\label{sigmaDH}
\end{equation}
and  the metric function $N(r)$ takes the form,
see Eq.~(\ref{N00}),
\begin{equation}
N(r) = 1-\frac{\Lambda
r^2}{3}+\frac{{Q_{m}^2}\left(1+\frac{2\Lambda
q}{3}\right)r^2-2Mr^3-
\frac{14}{5}{Q_{m}^2} q}{r^4+{2Q_{m}^2} q}
\,, \label{sNDH}
\end{equation}
At the center the metric function $N(r)$ takes the value
$N(0)=-2/5$.
Now we study the cases
$\Lambda>0$, $\Lambda=0$, $\Lambda<0$, and describe new fine
details of the horizon structure of the solutions
using the different
masses $M$ of the object.

\subsection{The case $\Lambda>0$}

\subsubsection{The distinguishing masses}

\begin{figure*}[t]
\halign{\hfil#\hfil&\hfil#\hfil\cr
\parbox[b]{0.5\textwidth}
{\centerline {\includegraphics[height=6cm]{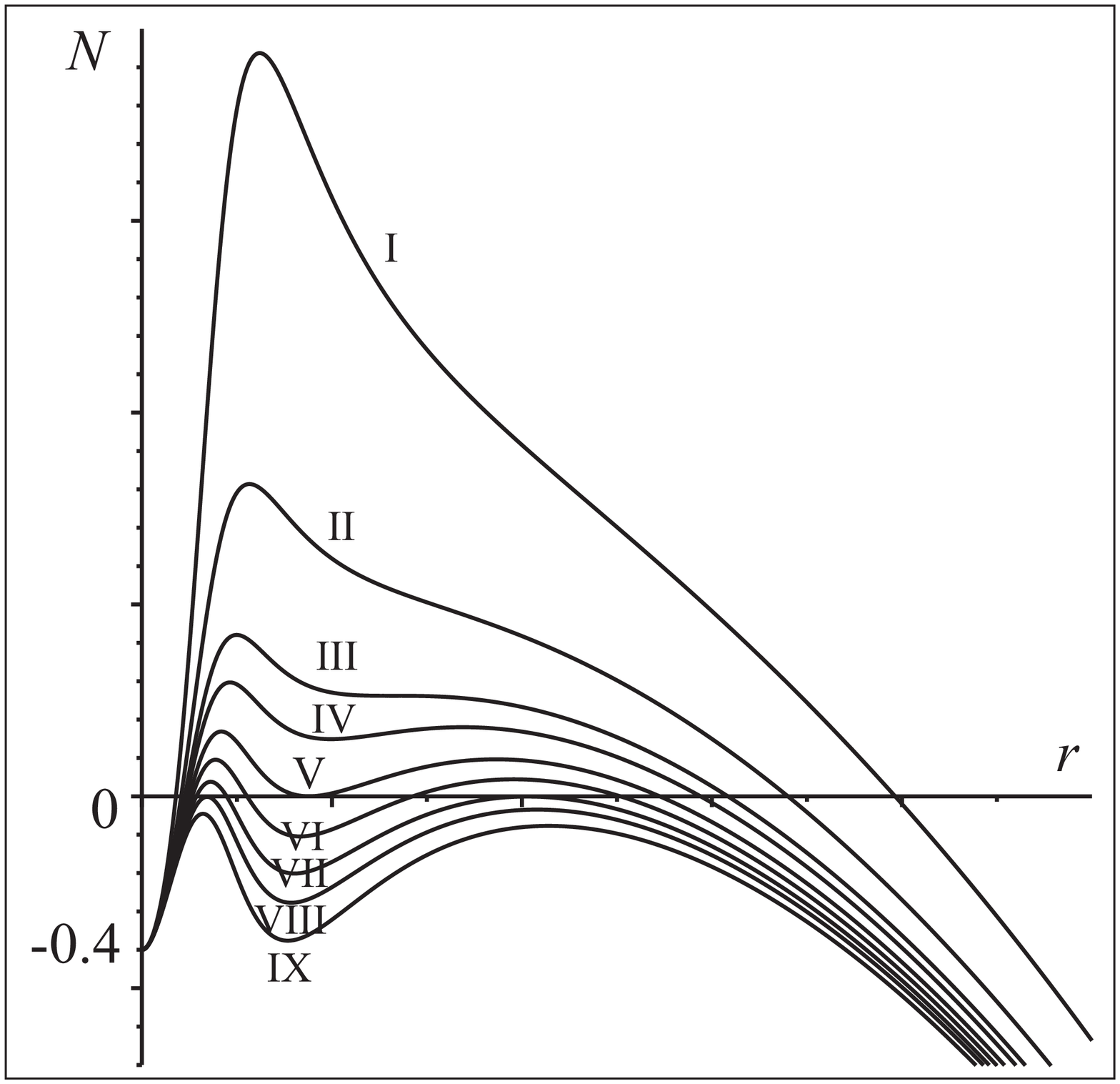}}
\centerline{(a)}\medbreak} & \parbox[b]{0.5\textwidth}
{\centerline{\includegraphics[height=5.2cm]{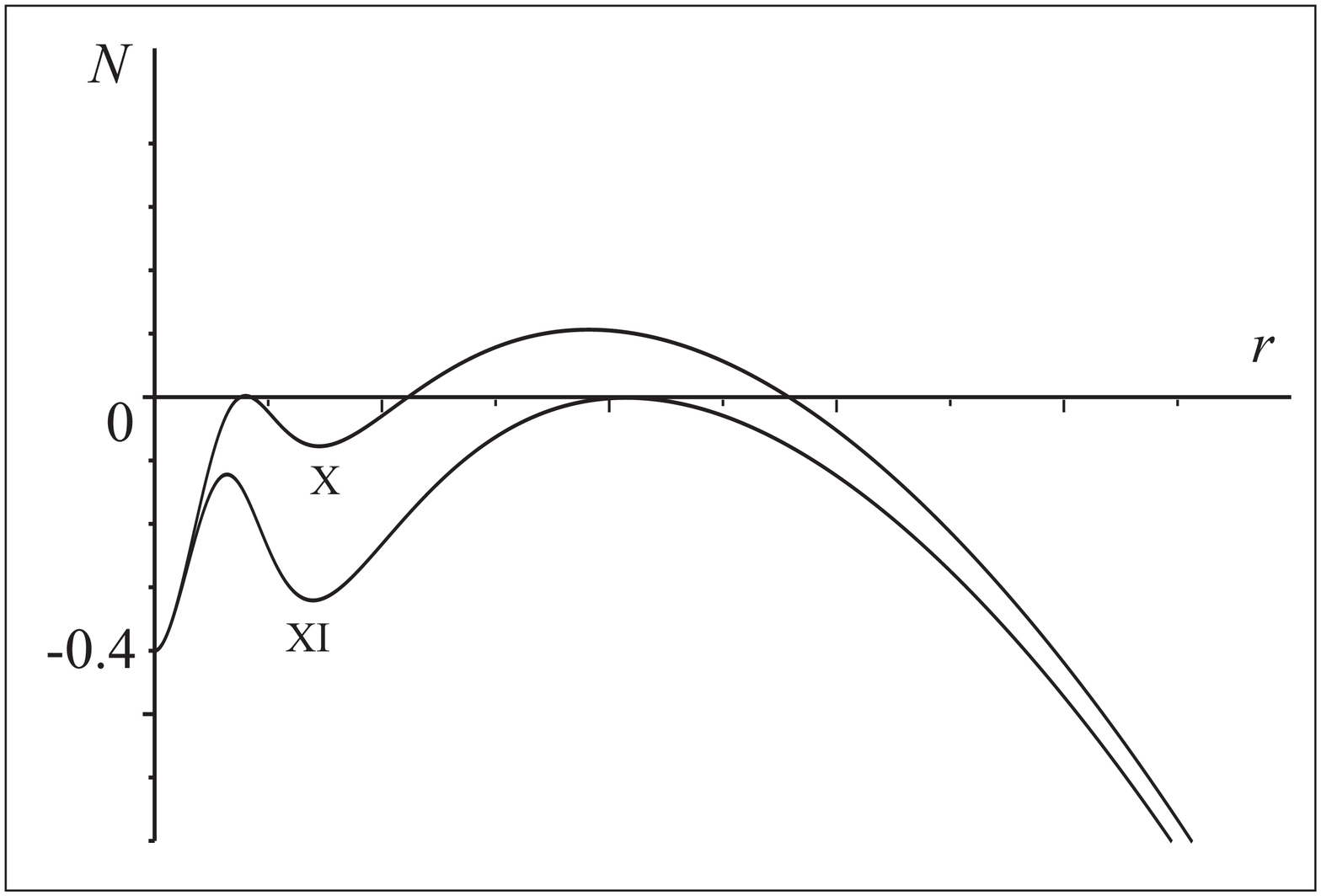}}
\centerline{(b)}\medbreak} \cr
\parbox[b]{0.5\textwidth}
{\centerline{\includegraphics[height=5.2cm]{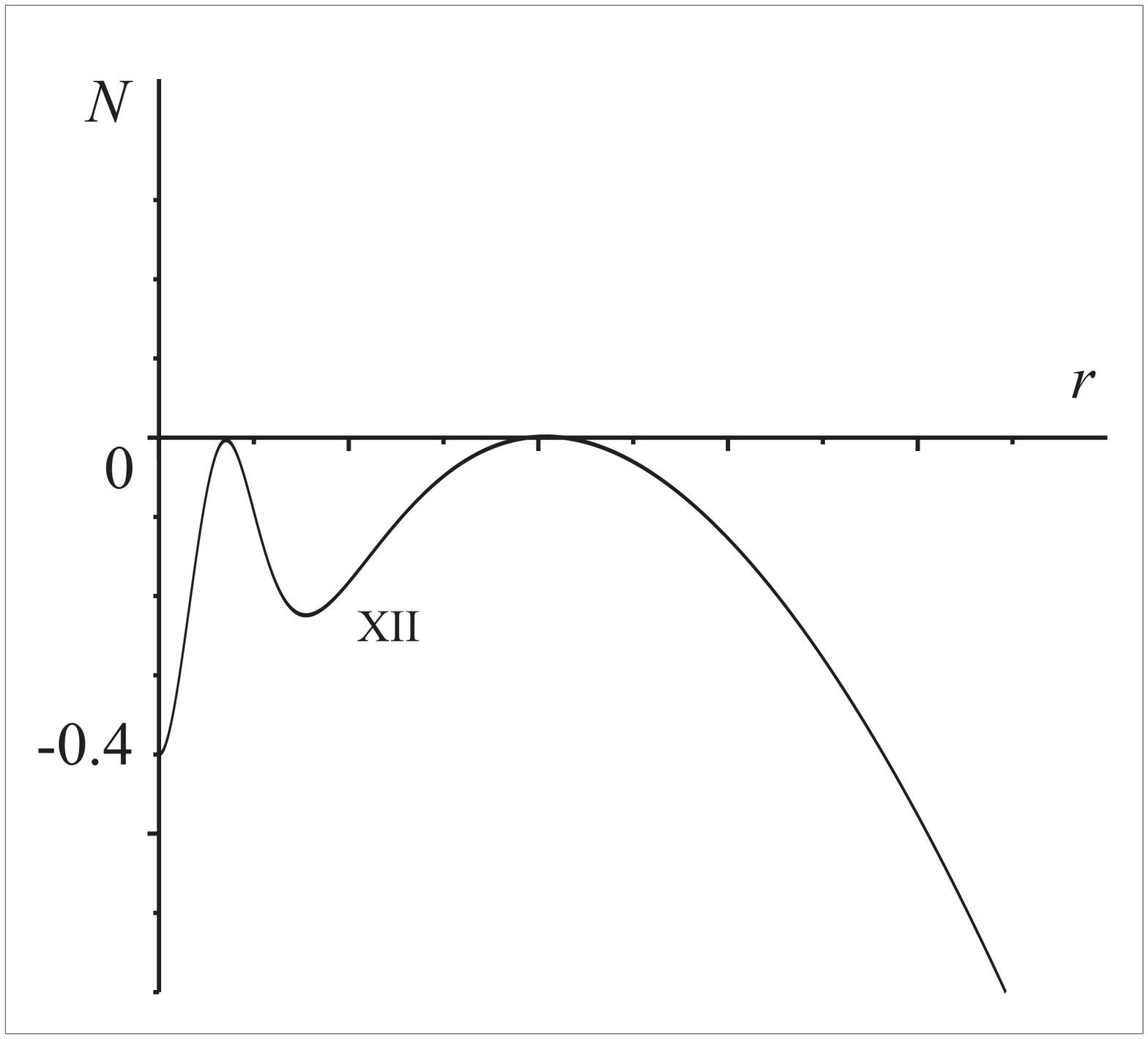}}
\centerline{(c)}\medbreak} & \parbox[b]{0.5\textwidth}
{\centerline{\includegraphics[height=5.2cm]{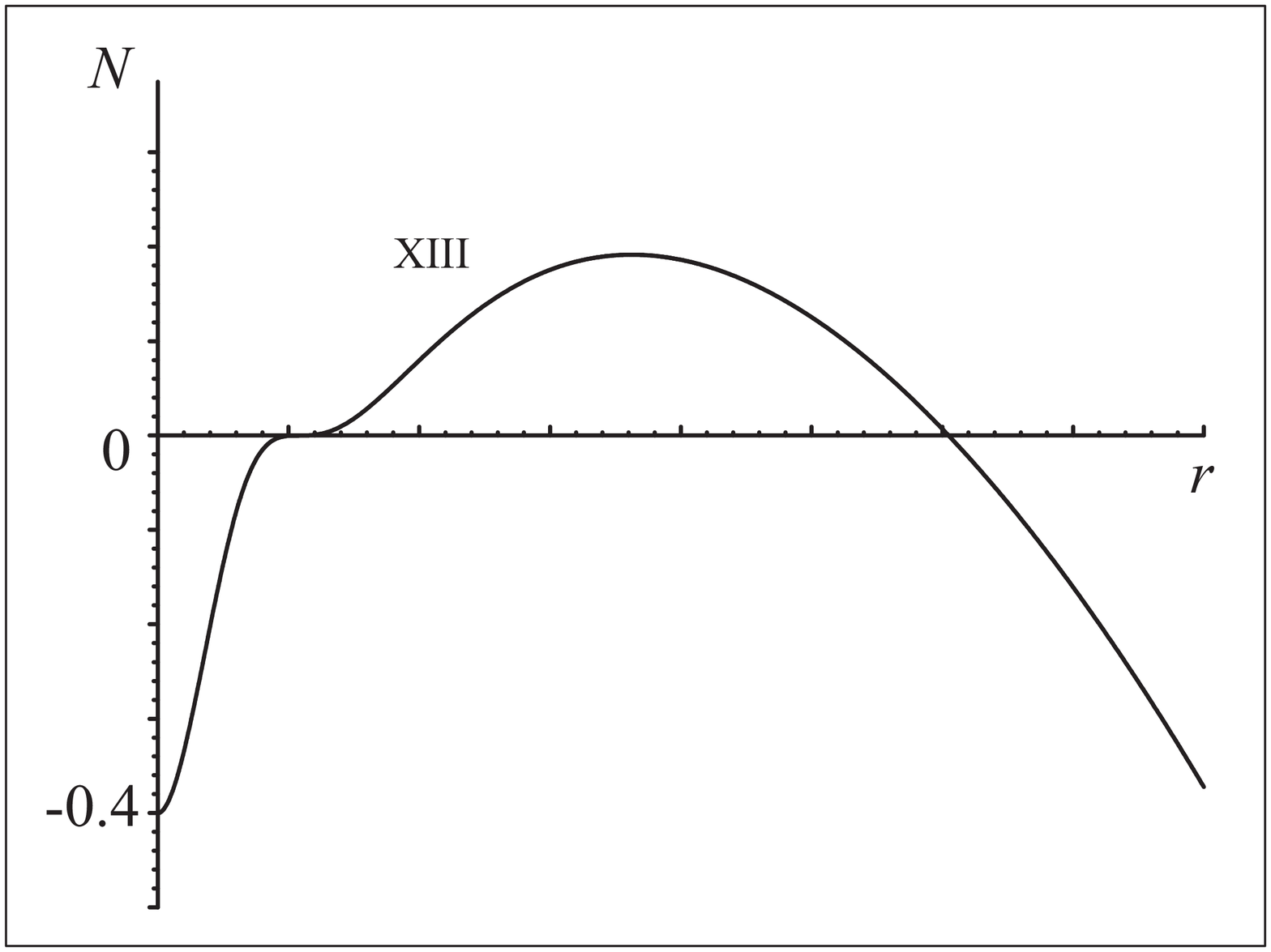}}
\centerline{(d)}\medbreak} \cr
\parbox[b]{0.5\textwidth}
{\centerline{\includegraphics[height=4.8cm]{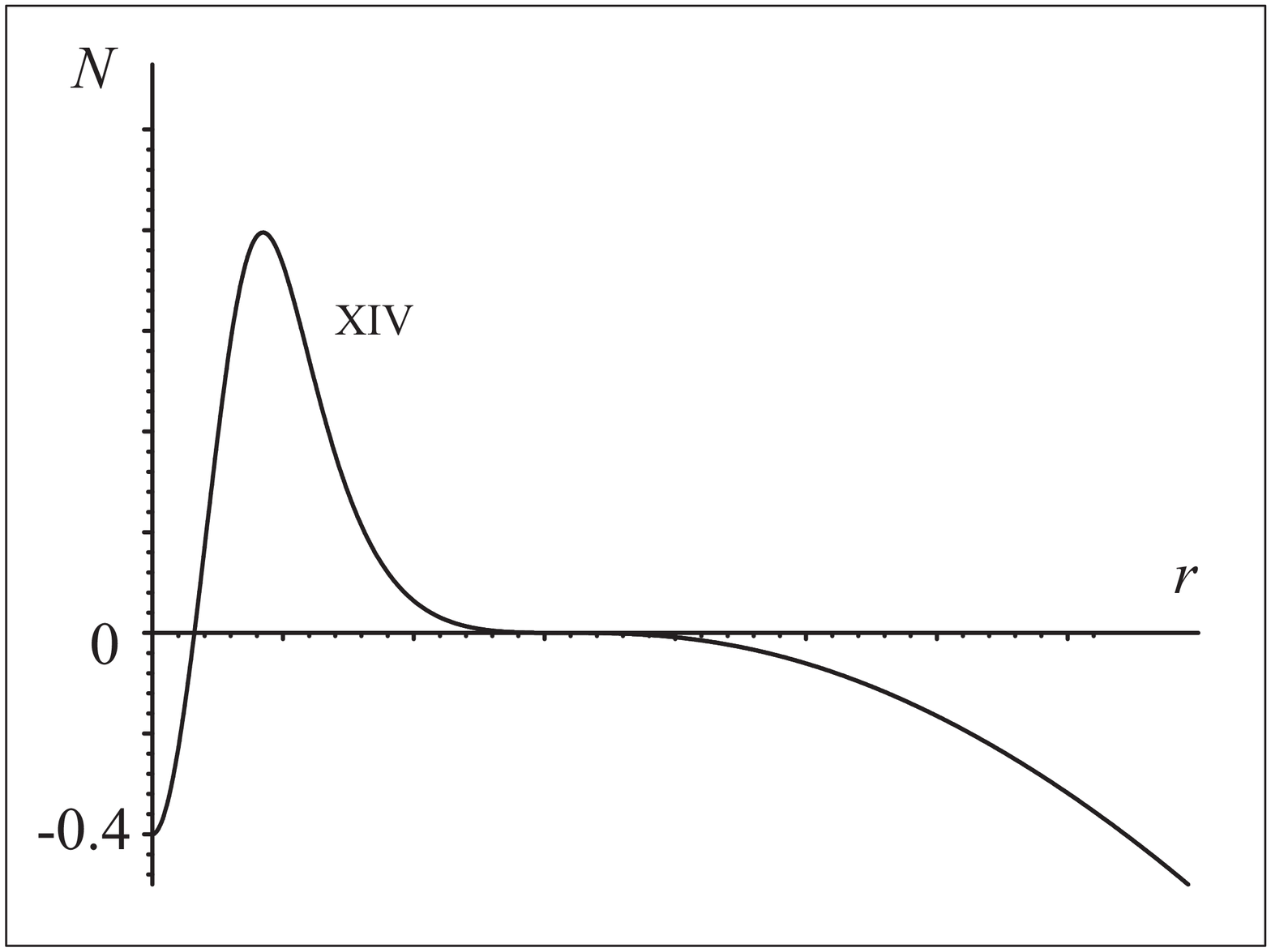}}
\centerline{(e)}\medbreak} & \parbox[b]{0.5\textwidth}
{\centerline{\includegraphics[height=5.2cm]{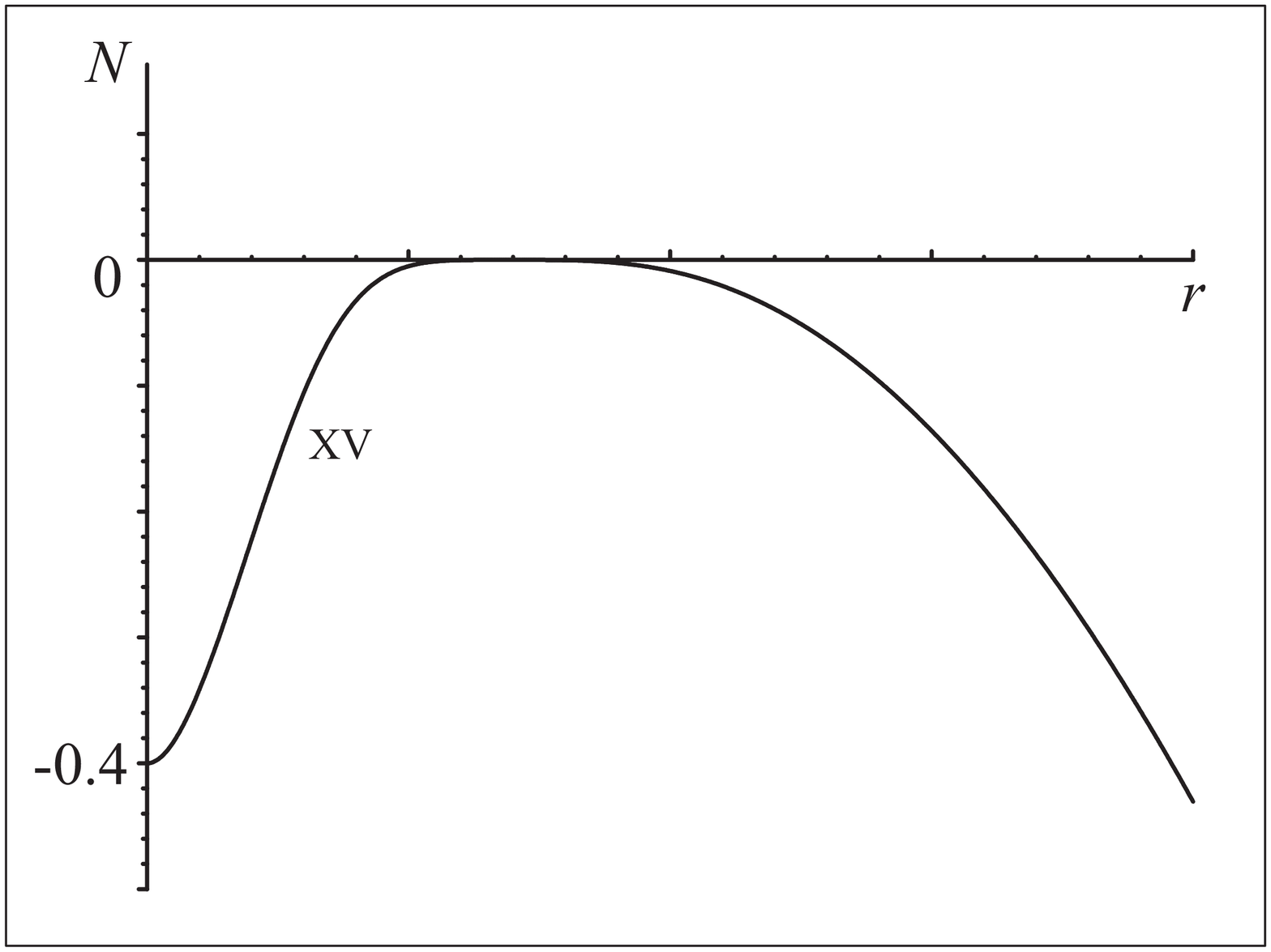}}
\centerline{(f)}\medbreak} \cr}
\caption{Quasiregular nonminimal black holes of the Drummond-Hathrell
type with $\Lambda>0$.
See text for details.
}\label{fig8}
\end{figure*}

For $\Lambda>0$  in the Drummond-Hathrell case the
corresponding figures are given in panels
$(1a)$ to $(1e)$ in Fig.~\ref{figc}.
We are faced with 15 different submodels,
see curves I-XV depicted in Fig.~\ref{fig8},
see also Table
I.

We distinguish these models with respect to the
mass $M$ of the object.
The analysis is rich but intricate
and is based on the introduction of the following
specific values of the mass.

\vskip 0.2cm
\noindent
$M_{{\rm N}}$:  We start with
the mass $M_{{\rm N}}$ that
distinguishes models with and without a Newtonian-type
attraction zone. For this value of the
mass the solutions of the two
equations, $N^{\prime}(r_{{\rm N}})=0$ and
$N^{\prime\prime}(r_{{\rm N}})=0$, coincide, but $N(r_{{\rm N}})
\neq 0$, see curve III in Fig.~\ref{fig8}.

\vskip 0.2cm
\noindent
$M_{1}$: The mass $M_{1}$ appears
when the minimum of the curve $N(r)$
touches the axis $N=0$, see curve V in Fig.~\ref{fig8}.
This means that the  two
equations, $N(r_{1})=0$ and $N^{\prime}(r_{1})=0$ give
coinciding solutions.

\vskip 0.2cm
\noindent
$M_{2}$: Similarly, the value $M_{2}$ is for the case
when the distant maximum touches the line $N=0$,
see curve VII on
Fig.~\ref{fig8}.

\vskip 0.2cm
\noindent
$M_{3}$:  Similarly, the value $M_{3}$
is for the case
when the maximum
closest to the center touches the line $N=0$,
see curve VIII  in Fig.~\ref{fig8}.

\vskip 0.2cm
\noindent
$M_{1m}$, $M_{2m}$, $M_{3m}$:
When the distant maximum is higher than the
maximum closest to the
center
the corresponding masses differ
from $M_{1}$, $M_{2}$, $M_{3}$.
In this case the
corresponding masses
are written as $M_{1m}$, $M_{2m}$, $M_{3m}$,
respectively, where the additional
index $m$, stands for modified.
In panel (b) of Fig.~\ref{fig8} other curves
could be drawn. We stick to drawing in this
panel (b) only the curves
that are qualitatively different from those
of panel (a). These are
curves X and XI only, which correspond to the masses $M_{2m}$ and
$M_{3m}$.  The curve for the mass value $M_{1m}$ is not displayed
since it can be obtained as a deformed curve V from panel (a) of
Fig.~\ref{fig8}.

\vskip 0.2cm
\noindent
$M_{2}=M_{3}$:
When $M_{2}=M_{3}$ we deal with a specific case
depicted in panel (c), curve XIII, of Fig.~\ref{fig8}.

\vskip 0.2cm
\noindent
$M_{{\rm T1}}$, $M_{{\rm T2}}$:
There are
also two specific sets of values of two parameters, the mass $M$
and  $\bar{q}$, for which three equations give the same roots,
namely, $N(r_{{\rm T}}) = 0$, $N^{\prime}(r_{{\rm T}})=0$ and
$N^{\prime\prime}(r_{{\rm T}})=0$, where
the subscript $\rm T$ is for triple. There
are two masses that fulfill these conditions.
One, $M_{{\rm
T1}}$, is depicted in
panel (d), curve XIV, of Fig.~\ref{fig8},
and shows the case when the inner, Cauchy,
and event horizons coincide.
The other, $M_{{\rm T2}}$, is depicted in
panel (e), curve XV, of Fig.~\ref{fig8},
and shows the case when the Cauchy,
event, and cosmological horizons coincide.

\vskip 0.2cm
\noindent
$M_{{\rm Q}}$:
The last specific value of the mass, $M_{{\rm Q}}$, appears
for a specific set $M_{{\rm Q}}$, $\bar{q}_{{\rm Q}}$,
$\Lambda_{{\rm Q}}$, for which $N(r_{{\rm Q}}) = 0$,
$N^{\prime}(r_{{\rm Q}})=0$, $N^{\prime\prime}(r_{{\rm Q}})=0$, and
$N^{\prime\prime\prime}(r_{{\rm Q}})=0$,
see the panel (f), curve XV, of Fig.~\ref{fig8}.

\subsubsection{The fifteen cases}

Now we comment the fifteen cases based on the
plots of $N(r)$ as a function of $r$.

\noindent
{\it (1) $M=0$ }

\noindent
This case is shown in curve I
of panel (a) of
Fig.~\ref{fig8} and is related to the
crossing of the line $(1a)$ on Fig.~\ref{figc}
with the line $M=0$, the bottom
horizontal line.
This solution has a similar structure
to the Schwarzschild-dS solution.
$N(r)$ is characterized by a finite minimum at the
center, an inner black hole zone and
an inner event horizon, a trap
between the horizon and the maximum, a zone of repulsion,
a cosmological horizon, and a dS region.

\noindent
{\it (2) $0<M<M_{{\rm N}}$}

\noindent
This case is shown in curve II of panel (a)
of Fig.~\ref{fig8} and looks
like the previous case $M=0$. The difference is in the height of the
maximum.

\noindent
{\it (3) $M=M_{\rm N}$}

\noindent
This case is shown in curve III of panel (a) of Fig.~\ref{fig8}.
It separates spacetimes with and without a Newtonian-type attraction
zone. This solution has also a similar structure to the
Schwarzschild-dS solution.  $N(r)$ is characterized by a
minimum at the center, an inner black hole zone, an inner event
horizon, a trap between the horizon and the maximum, a zone of
repulsion, a cosmological horizon, and a dS region.

\noindent
{\it (4) $M_{{\rm N}}<M<M_{1}$}

\noindent
This case is shown in curve IV  of panel (a) of
Fig.~\ref{fig8}. The curve
is characterized by a minimum at the center, an
inner black hole zone,
an inner event horizon, a trap between the horizon and the maximum,
a zone of repulsion, a neutral zone, a Newtonian attraction zone,
 a zone of
repulsion, a cosmological horizon, and a dS region.

\noindent
{\it (5) $M=M_{1}$}

\noindent
This case is shown in curve V of panel (a)
of Fig.~\ref{fig8}.  There are
three horizons, one of them is a double horizon.  It can be thought of
as extremal Reissner-Nordstr\"om-dS solution, with a
Schwarzschild structure replacing the timelike singularity.  The curve
is characterized by a minimum at the center, an inner black hole zone,
an inner black hole horizon, a trap between the horizon and the
maximum, a zone of repulsion, a double extremal horizon, a Newtonian
attraction zone, a zone of cosmological repulsion, a cosmological
horizon, and a dS region.  This case corresponds to the cross
of the curve in panel $(1a)$ of Fig.~\ref{figc}, for which the
horizontal mass line touches the minimum of the plot of the auxiliary
function $f(r)$.

\noindent
{\it (6) $M_{1}<M<M_{2}$ }

\noindent
This case is shown  in curve VI of panel (a) of
Fig.~\ref{fig8}.  There are four horizons. It can be thought of as a
Reissner-Nordstr\"om-dS solution, with a Schwarzschild
structure replacing the timelike singularity.  $N(r)$ is characterized
by a minimum at the center, an inner black hole zone, an inner event
horizon, a trap between the horizon and the maximum, a zone of
repulsion, a Cauchy horizon, a second black hole zone, a second event
horizon, a Newtonian attraction zone, a zone of cosmological
repulsion, and a cosmological horizon, and a dS region.  This
case corresponds to the situation, for which the horizontal mass line
crosses the plot of the auxiliary function in four different points,
see panel $(1a)$ of Fig.~\ref{figc}.

\noindent
{\it (7) $M=M_{2}$ }

\noindent
This case is shown in curve VII of panel (a) of
Fig.~\ref{fig8}.  There are
three horizons, one of them is a double horizon.
It can be thought of
as Reissner-Nordstr\"om-dS solution, where
the second event horizon coincides with the cosmological
horizon, and
with a
Schwarzschild structure replacing the timelike singularity.
The
curve is characterized by a minimum at the center, an inner black hole
zone, an inner black hole horizon, a small trap between the horizon and the
maximum, a zone of repulsion, a Cauchy horizon, a second black hole zone,
a double horizon where the  second event horizon
coincides with the cosmological
horizon, and a dS region.
This case corresponds to the situation for which the
horizontal mass line touches the plot of the auxiliary function in the
right maximum see panel $(1a)$ of Fig.~\ref{figc}.

\noindent
{\it (8) $M=M_{3}$ }

\noindent
This case is shown in curve VIII of Fig.~\ref{fig8}.
This solution can be thought of
as a limiting Kasner type solution
$N(r)$ is always negative, apart at one point where
it is zero.
This case corresponds to
the situation, for which the horizontal mass line touches the plot of
the auxiliary function in the left maximum,
see panel $(1a)$ of Fig.~\ref{figc}.

\noindent
{\it (9) $M>M_{3}$ }

\noindent
This case is shown in curve IX of Fig.~\ref{fig8}.
This solution can be thought of
as a Kasner type solution
$N(r)$ is always negative.
There are no horizon.
This case corresponds to the situation, for which the
horizontal mass line is situated above the plot of the auxiliary
function,
see panel $(1a)$ of Fig.~\ref{figc}.

\noindent
{\it (10) $M=M_{3m}$ }

\noindent
This case is  shown in curve X of
panel (b) of Fig.~\ref{fig8}.
It is a new case.
There are three horizons, one is a double horizon,
the inner event horizon and the Cauchy horizon
coincide.
It can be thought as having the same structure
of the
Reissner-Nordstr\"om-dS solution.
For this case the curve
has a
minimum at the center, an
inner black hole Reissner-Nordstr\"om
zone, a double horizon where the inner event horizon and the Cauchy horizon
coincide, a second black hole zone, outer horizon, a Newtonian
attraction zone, a zone of cosmological repulsion,
a cosmological horizon, and a dS region.

 \noindent
{\it (11) $M=M_{2m}$ }

\noindent
This case is shown in curve XI of
panel (b) of Fig.~\ref{fig8}.
It is a limiting Kasner type solution.

\noindent
{\it (12) $M=M_{2m}=M_{3m}$}

\noindent
This case is shown in curve XII of
panel (c) of Fig.~\ref{fig8}.
It is also a limiting Kasner type solution.

\noindent
{\it (13) $M=M_{{\rm T1}}$}

\noindent
This case is shown in curve XIII of
panel (d) of Fig.~\ref{fig8}.
This solution has also a similar structure
to the Schwarzschild-dS solution.
For this case there is
a minimum at the center, a triple
inner, Cauchy and outer horizons, a
Newtonian attraction zone, a
zone of cosmological repulsion, a cosmological  horizon,
and a dS region.

\noindent
{\it (14)  $M=M_{{\rm T2}}$}

\noindent
This case is shown in curve XIV of
panel (e) of Fig.~\ref{fig8}.
This solution has also a similar structure
to the Schwarzschild-dS solution.
For this case there is a minimum at the center, an inner black hole
zone, an inner event horizon, a trap, a zone of repulsion,
triple horizon where the
Cauchy, the second event, and
the cosmological horizons coincide, and a dS region.
This situation corresponds to
the case, when the horizontal mass line crosses the plot of the
auxiliary function in the triple point depicted on panel $(1c)$ of
Fig.1.

\noindent
{\it (15) $M=M_{{\rm Q}}$}

\noindent
This case is shown in curve XV of
panel (e) of Fig.~\ref{fig8}.
It is also a limiting Kasner type solution.
This situation corresponds to the case, when the horizontal mass line
crosses the plot of the auxiliary function in the quadruple point
depicted on panel $(1d)$ of Fig.1.

With the features given above
one can sketch with some ease the corresponding
Carter-Penrose diagrams.

\subsection{The case $\Lambda=0$}

\begin{figure*}[t]
\halign{\hfil#\hfil&\hfil#\hfil\cr
\parbox[b]{0.5\textwidth}{\centerline
{\includegraphics[height=5.2cm]{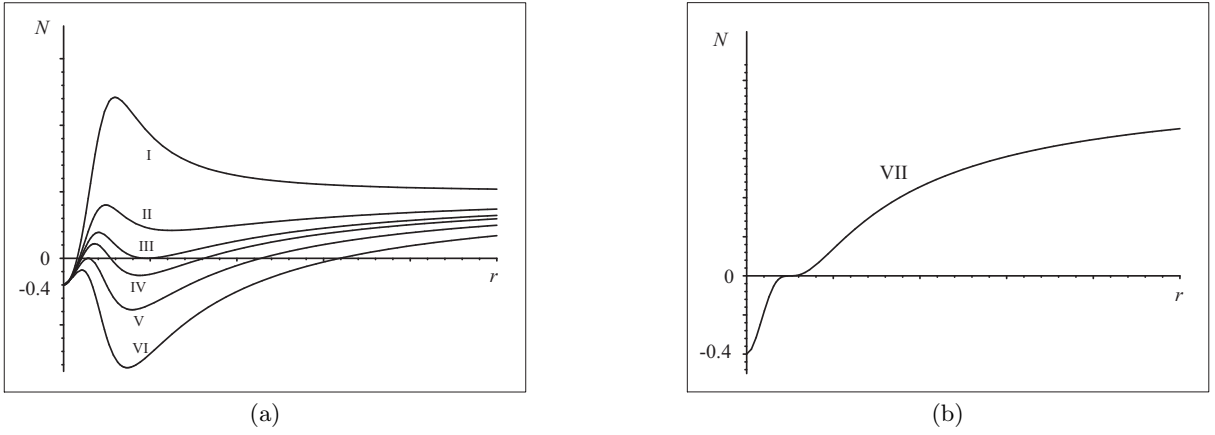}}
\centerline{(a)}\medbreak} & \parbox[b]{0.5\textwidth}
{\centerline{\includegraphics[height=5.2cm]{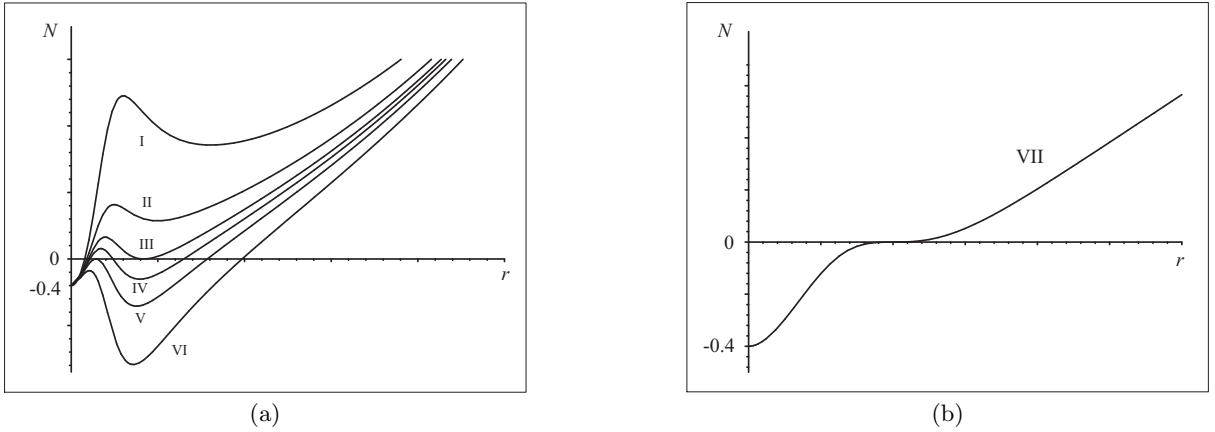}} \centerline{(b)}
\medbreak} \cr}
\caption{Quasi-regular nonminimal black holes of the Drummond-Hathrell
type with $\Lambda=0$. See text for details.}\label{fig6}
\end{figure*}

\begin{figure*}[t]
\halign{\hfil#\hfil&\hfil#\hfil\cr
\parbox[b]{0.5\textwidth}{\centerline{\includegraphics[height=5.2cm]{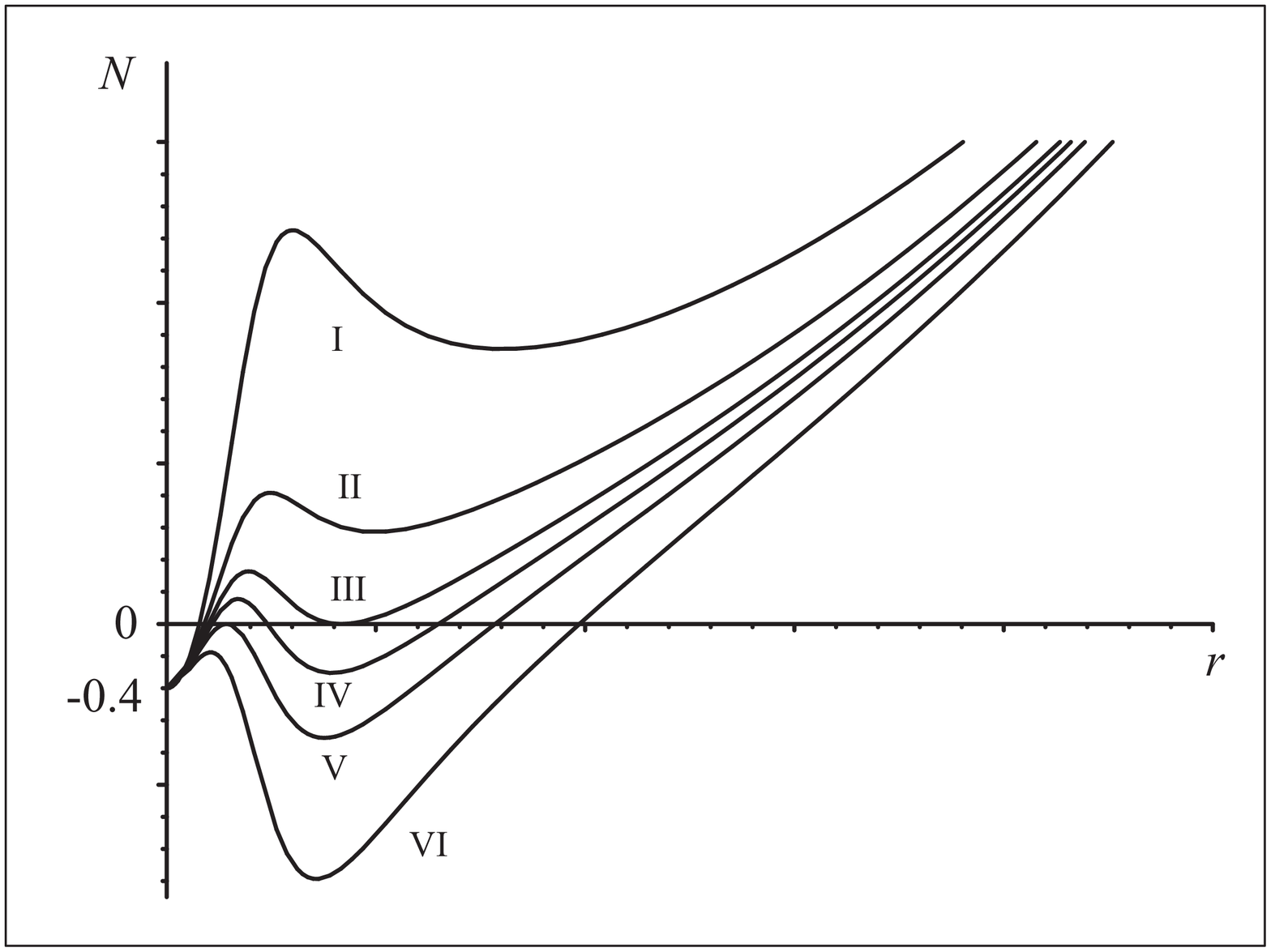}}
\centerline{(a)}\medbreak} & \parbox[b]{0.5\textwidth}
{\centerline{\includegraphics[height=5.2cm]{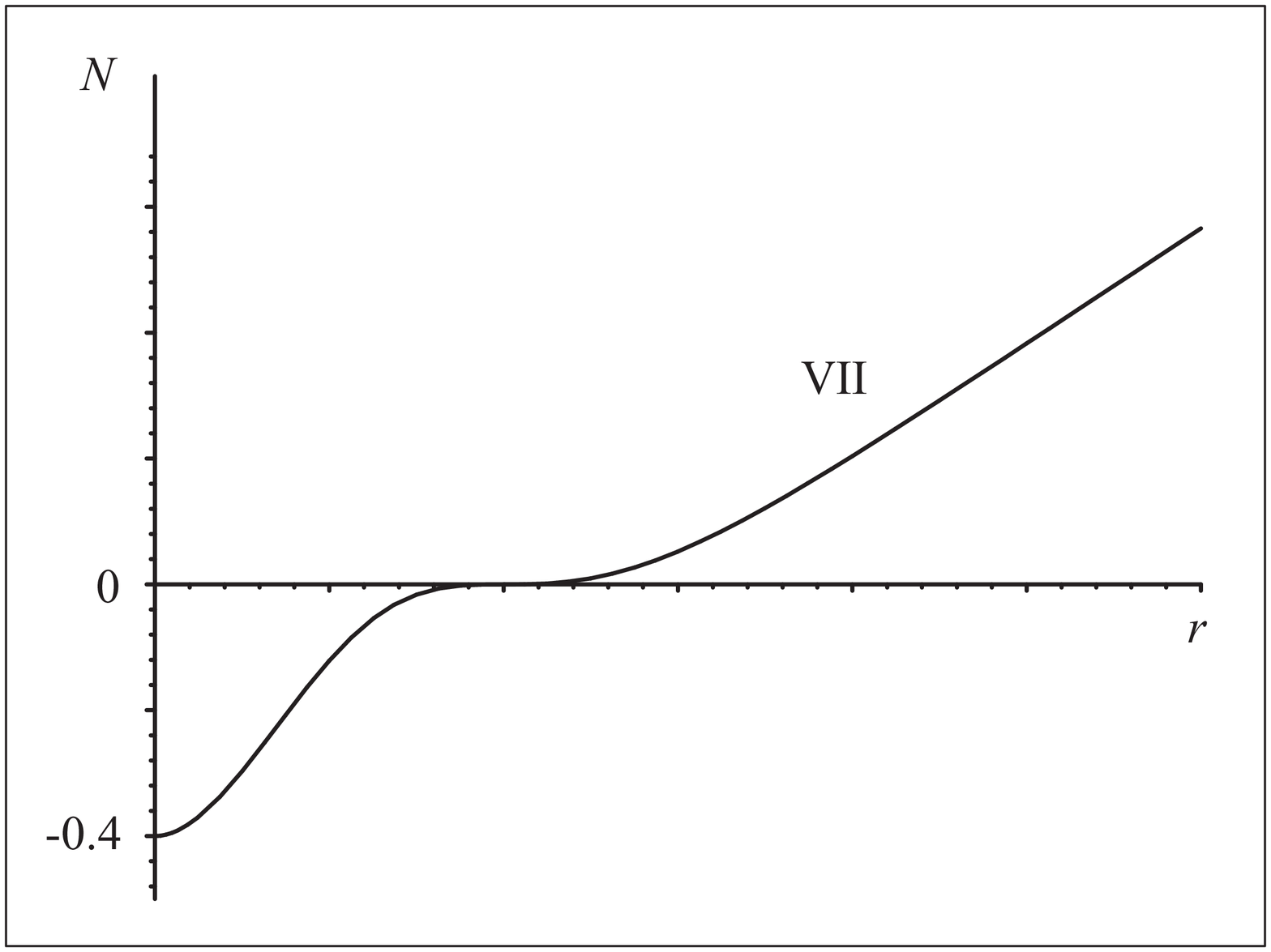}}
\centerline{(b)}\medbreak} \cr}
\caption{Quasi-regular nonminimal black holes of the
Drummond-Hathrell type with $\Lambda<0$. See text for details.
}\label{fig7}
\end{figure*}

\subsubsection{The distinguishing masses}

For $\Lambda=0$  in the Drummond-Hathrell case the
corresponding figures are given in panels
$(1a)$ to $(1c)$ in Fig.~\ref{figb}.
We have 7 different submodels,
see curves I-VII depicted in Fig.~\ref{fig6},
see also Table I.

For $\Lambda=0$
the spacetime is
asymptotically Minkowskian, i.e., $N(r) \to
1$ at $r\to \infty$.
All the
curves have a minimum at the center with $N(0)=-\frac25$, an inner
black hole zone, and an inner event horizon.

There are two
distinguishing masses, $M_1$ and $M_2$.
The mass $M_1$ is related to the case when curve III on panel (a) of
Fig.~\ref{fig6}
touches the axis $N=0$ in its minimum. The mass $M_2$
corresponds to the case when curve V  of
Fig.~\ref{fig6}
touches this axis in its
maximum. There are seven intrinsic cases distinguished according to
different values of the mass $M$.

\subsubsection{The seven cases}

\noindent
{\it (1) $M=0$ }

\noindent
This case is shown in curve I
of panel (a) of
Fig.~\ref{fig6} and is related to the
crossing of the line (2a) on Fig.~2
with the line $M=0$, the bottom
horizontal line.
$N(r)$ is characterized by a finite minimum at the
center, an inner black hole zone,
an inner event horizon, a trap
between the horizon and the maximum, a zone of repulsion
and an asymptotic flat region.
This case corresponds to the cross
of the curve in panel (2a) of Fig.~\ref{figb}, for which the
bottom horizontal mass line crosses $f(r)$.

\noindent
{\it (2) $0<M<M_{1}$}

\noindent
This case is shown in curve II  of panel (a) of
Fig.~\ref{fig6}. The curve
is characterized by a minimum at the center, an
inner black hole zone,
an inner event horizon, a trap between the horizon and the maximum,
a zone of repulsion, a neutral zone, a Newtonian attraction zone,
and an asymptotic flat region.
As in the previous case, this case corresponds to the cross
of the curve in panel (2a) of Fig.~\ref{figb}, for which the
bottom horizontal mass line crosses $f(r)$.

\noindent
{\it (3) $M=M_{1}$}

\noindent
This case is shown in curve III of panel (a)
of Fig.~\ref{fig6}.   There are
two horizons, one of them is a double horizon.  It can be thought of
as extremal Reissner-Nordstr\"om solution, with a
Schwarzschild structure replacing the Reissner-Nordstr\"om
timelike singularity.  The curve
is characterized by a minimum at the center, an inner black hole zone,
an inner black hole horizon, a trap between the horizon and the
maximum, a zone of repulsion, a double extremal horizon, a Newtonian
attraction zone,
and an asymptotic flat region.
This case corresponds to the cross
of the curve in panel (2a) of Fig.~\ref{figb}, for which the
horizontal mass line touches the minimum of the plot of the auxiliary
function $f(r)$.

\noindent
{\it (4) $M_{1}<M<M_{2}$ }

\noindent
This case is shown  in curve IV of panel (a) of
Fig.~\ref{fig6}.  There are three horizons. It can be thought of as a
Reissner-Nordstr\"om solution, with a Schwarzschild
structure replacing the Reissner-Nordstr\"om
timelike singularity.  $N(r)$ is characterized
by a minimum at the center, an inner black hole zone, an inner event
horizon, a trap between the horizon and the maximum, a zone of
repulsion, a Cauchy horizon, a second black hole zone, a second event
horizon, a Newtonian attraction zone,
and an asymptotic flat region.  This
case corresponds to the situation, for which the horizontal mass line
crosses the plot of the auxiliary function in three different points,
see panel (2a) of  Fig.~\ref{figb}.

\noindent
{\it (5) $M=M_{2}$ }

\noindent
This case is shown in curve V of panel (a) Fig.~\ref{fig6}.
There are
two horizons, one of them is double.  The curve is characterized by a
minimum at the center, an inner black hole zone, a double horizon, in
which an inner and a Cauchy horizons coincide, a second black hole
zone, a second event horizon, a Newtonian attraction zone,
and an asymptotic flat region.  This case
corresponds to the situation for which the horizontal mass line
touches the plot of the auxiliary function in the maximum see the
panel (2a) of  Fig.~\ref{figb}.

\noindent
{\it (6) $M>M_{2}$ }

\noindent
This case is shown in curve VI of Fig.~\ref{fig6}.  The curve is
characterized by a minimum at the center, a united black hole zone,
an event horizon, a Newtonian attraction zone,
and an asymptotic flat region. This case corresponds to
the situation for which the horizontal mass line is situated above
the maximum of the auxiliary function, see panel (2a) of Fig.~\ref{figb},
or
crosses the line depicted on panel (2b) of Fig.~\ref{figb}.

\noindent
{\it (7) $M=M_{1}=M_{2}$}

\noindent
This case is shown in curve VII of
panel (b) of Fig.~\ref{fig6}.
One has one triple horizon, where the inner,
Cauchy, and event horizons coincide.
The
curve is characterized by a minimum at the center,
an inner black hole zone, a triple horizon,
a Newtonian attraction zone,
and an asymptotic flat region. This case
corresponds to the situation for which the
horizontal mass line crosses the plot of the auxiliary
function in the triple point, see panel (2c) of Fig.~\ref{figb}.

With the features given above
one can sketch with some ease the corresponding
Carter-Penrose diagrams.

\subsection{The case $\Lambda<0$}

\subsubsection{The distinguishing masses}

From the point of view of horizon structure and
description, the case with negative
cosmological constant, $\Lambda<0$,
does not differ qualitatively from the case
$\Lambda=0$.
There are two masses $M_1$ and $M_2$ as in the $\Lambda=0$
case.

\subsubsection{The seven cases}

We draw seven subcases illustrated by
curves I-VI on panel (a) of  Fig.~\ref{fig7}
and by curve VII on panel
(b) of  Fig.~\ref{fig7}. The main difference
to the $\Lambda=0$ case
is that all the curves are asymptotically
anti-de
Sitter instead of asymptotically flat.
The details are
similar to the ones for the case $\Lambda=0$.

With the features given above
one can sketch with some ease the corresponding
Carter-Penrose diagrams.

\section{The regular black hole:
Example of a
nonminimal theory with $\bar q\geq0$}
\label{secreg}

If we want regular solutions at the center then we have to
impose further
$N(0)=1$ and $N^{\prime}(0)=0$. From Eq.~(\ref{N002})
we see this happens when
\begin{equation}
\bar q=q \,. \label{barqq}
\end{equation}
Since we assume $q>0$, see Eq.~(\ref{q>0}),
the regular solutions fall in the
case $\bar q \geq0$, the case $\bar q=0$
being a limiting case.
In addition, the requirement given in
Eq.~(\ref{barqq}),
restrict the number of the two nonminimal
coupling constants, $q$ and $\bar q$, to just one
independent coupling constant, $q$, say.
We deal with a four parameter model.

Then the solution to Eq.~(\ref{N00}) is of the form
\begin{equation}
N= 1 +\left(1+\frac{{2Q_{m}^2} q}{r^4}\right)^{-1}\left(-\frac{2M}{r}
+ \frac{{Q_{m}^2}}{r^2} -\frac{\Lambda}{3}r^2 \right)\,.
\label{N00xx}
\end{equation}
The four parameters in this  family of exact solution
are them
$q$, $M$, $Q_m$, and $\Lambda$.
Near the
center the metric function $N(r)$ behaves as
$
N(r) =  1 + \frac{r^2}{2 q}  - \frac{Mr^3}{{Q_{m}^2} q} +
\dots$, such that
$N(0)=1$, $N^{\prime}(0)=0$ and $N^{\prime
\prime}(0)= \frac{1}{q}$. This means that the point $r=0$ is a
minimum of the regular function $N(r)$ independently of the sign and
value of the cosmological constant $\Lambda$, and independently of
the mass value $M$. Since $N(0)=1$, the curvature scalar is
regular in the center: $R(0)=\frac{6}{q}$. The quadratic scalar
$R_{mn}R^{mn}= \frac{9}{q^2}$, and other curvature invariants are
also finite in the center. Thus the spacetime is regular at
the center.
All the corresponding objects found within this solution
are regular objects.
The most important feature of the family of exact solutions is that it
has solutions with horizons, i.e.,
regular black holes, depending on the relative values of the
parameters.
These solutions have been displayed in detail
in \cite{balaklz2}, below we give a brief account of them.

\subsection{The case $\Lambda>0$}

For $\Lambda>0$, there are two critical masses
$M_{\rm c1}$ and $M_{\rm c2}$.
Depending on the values of the parameters, the black
hole solution can have three horizons, the Cauchy horizon, the event
horizon and the cosmological horizon.  When $M<M_{\rm c1}$,
there is one horizon only which is a cosmological
horizon.  A typical profile of the metric function
$N(r)$ contains a central small cavity, a repulsion barrier, a zone of
rest  near the point of minimum, a Newtonian-type attraction
zone with the potential going as $1/r$, a cosmological
acceleration zone, and a asympotically dS zone.
When $M=M_{\rm c1}$, there is an extremal
horizon that is double,
formed by the Cauchy and event horizons, and there is
a cosmological horizon. When $M_{\rm c1}< M<M_{\rm c2}$ there are
three separate horizons, the Cauchy, event and cosmological
horizons.  When $M=M_{\rm c2}$, there is a Cauchy horizon,
and there is a double horizon, the event horizon and the
cosmological horizon coincide.  In this case the black hole is a
cosmological supermassive regular  extremal
black hole. The whole
visible universe is swallowed by this supermassive object.  For $
M>M_{\rm c2}$ the Cauchy horizon becomes a cosmological horizon, the
black hole is ultramassive. In such a universe there is only one
horizon, which is cosmological, together with a repulsion region.
This ultramassive black hole is of a new type,
the three horizons coincide: the Cauchy, event and the cosmological
horizons.  There is further
the case in which the three horizons coincide
when $M_{\rm c1}=M_{\rm c2}$.
With the features given above
one can sketch the corresponding
Carter-Penrose diagrams.
For further details see \cite{balaklz2}.

\subsection{The case $\Lambda=0$}

For $\Lambda =0$ spacetime is asymptotically flat.
Depending on the parameters, the black hole solution can have two
horizons, the Cauchy horizon and the event horizon. When the mass is
below a certain critical mass $M_{\rm c}$ there are no horizons.
There is a double horizon when $M=M_{\rm
c}$, and when  $M>M_{\rm c}$ the
Cauchy horizon and event horizons stand alone.
With the features given above
one can sketch the corresponding
Carter-Penrose diagrams.
For further details see \cite{balaklz2}.

\subsection{The case $\Lambda<0$}

For $\Lambda <0$ spacetime is asymptotically dS, and there is
no cosmological horizon. Depending on the values of the parameters,
the solution can have the Cauchy horizon and the event horizons. The
critical mass $M_{\rm c}$ is a mass below which there are stars and
above which there are regular black holes with the two horizons.  When
$M=M_{\rm c}$ the black hole has a double extremal horizon.
With the features given above
one can sketch the corresponding
Carter-Penrose diagrams.
For further details see \cite{balaklz2}.

\section{Conclusions}\label{Disc}

In this work we have found a general
exact spacetime solution for a
Wu-Yang magnetic monopole in
a nonminimal Einstein-Yang-Mills
theory. This general solution is in fact
a family of solutions with six parameters
and they generically represent
objects that go from bare
magnetic monopoles to black holes with magnetic
charge and several different types of horizons.

By a judicious choice we have reduced the number
of parameters of the solution from six to five.
Indeed imposing that
the singularities at the center are
spherical conical singularities, we have reduced
the number of nonminimal parameters from three to
two, $q$ and $\bar q$. The other three parameters
are the cosmological constant $\Lambda$,
the Wu-Yang magnetic charge represented by $Q_{m}$,
and the mass $M$.

We have provided a complete classification of these families of
magnetic monopole solutions with respect to the number of horizons and
their type.  The important parameters in this classification are
$\bar{q}$, $Q_{m}$, and $\Lambda$, together with the mass $M$ of the
spacetime. These furnish if there are zero, one, two, three, or four
horizons, and whether they have a simple, double, triple, or quadruple
degeneracy. The distinct horizons that appear within these families
of objects are inner, Cauchy, event horizons, as well as
a cosmological horizon
when $\Lambda$ is positive.

The objects have a great deal of unsuspected structure.
They have a trapping parabolic region near the center
controlled by the nonminimal parameters $q$ and
$\bar q$.
The
point $r=0$ is an equilibrium point for which $N^{\prime}(0)=0$.
When $\bar q>0$, $N(0)>0$, while for
$\bar q<0$ one has $N(0)>0$, and so in this case
the there is an inner horizon at a small positive $r$.
For $\bar q=0$ the horizon is at $r=0$, and so the
horizon and the conical singularity mesh in a null
singular horizon.
There is then a
repulsion barrier
contiguous to the nonminimal trap. Cauchy horizon
and event horizons can then also appear, and
in the positive cosmological constant case
a cosmological dS horizon appears.

These general features of the families of exact
solutions have been worked out
for two special cases, the Drummond-Hathrell model
and the regular black holes, examples of $\bar{q}<0$ and
$\bar{q}>0$, respectively.

It will be certainly interesting to find magnetic
monopole black hole solutions with an ansatz
different from the Wu-Yang ansatz. These
solutions would give nonminimal black holes
and monopoles with Yang-Mills hair.

\begin{acknowledgments}
ABB and AEZ thank financial support from the Program of Competitive
Growth of Kazan Federal University (KFU)
Project No.~0615/06.15.02302.034 and from the Russian
Foundation for Basic Research Grant (RFBR) No.~14-02-00598.  ABB
acknowledges financial support provided under the European
Union's H2020 ERC Consolidator Grant ``Matter and
strong-field gravity: New frontiers in Einstein's theory'',
Grant No.~MaGRaTh–646597.
JPSL thanks Funda\c c\~ao para a Ci\^encia e Tecnologia
(FCT) - Portugal for financial support through Project
No.~PEst-OE/FIS/UI0099/2014.
\end{acknowledgments}

\end{document}